# Complex crystal structure prediction using ML-enhanced multi-minima iterative genetic algorithm

Ling Tang[1], Weiyi Xia[2,3], Tyler J. Slade[2,3], Paul C. Canfield[2,3] and Cai-Zhuang Wang[2,3,*]

[1]School of Physics and Optical Engineering, Zhejiang University of Technology, Hangzhou, 310023, China
[2]Ames National Laboratory, U.S. Department of Energy, Ames, IA 50011, United States.
[3]Department of Physics and Astronomy, Iowa State University, Ames, IA 50011, United States

Corresponding author: *wangcz@ameslab.gov

## Abstract

Current machine learning (ML) approaches for materials discovery rely heavily on known structural databases, limiting their ability to identify entirely novel structure types. In this work, we develop a multi-minima iterative genetic algorithm (MMIGA) that integrates an artificial-neural-network machine learning (ANN-ML) interatomic potential with an iterative, metadynamics-inspired penalty scheme. We demonstrate the robustness of this method on a complex ternary La-Co-Pb system, characterized by Co-Pb immiscibility and an intricate energy landscape. The ML-enhanced MMIGA successfully predicts the ground-state *Pbam* structure of the recently synthesized $La_4Co_4Pb$ antagonistic-pair-phase, a novel structure missed by previous database-reliant ML predictions, while also identifying multiple metastable competing phases. Additionally, we challenged the MMIGA method to predict the structure of $La_5CoPb_2$ antagonistic-pair-phase, a new compound discovered during earlier attempts to synthesize the predicted phase $La_3CoPb$. With only knowledge of the composition, our MMIGA approach successfully predicts the orthorhombic structure of $La_5CoPb_2$, producing an exact match with the structure independently determined by x-ray diffraction. By efficiently mapping both global minimum and relevant competing metastable states, this approach provides critical theoretical insights into phase selection for novel quantum and magnetic materials.

## 1. Introduction

Novel advanced materials discovery and development is critical for driving technological innovation and sustainability. The vast compositional and structural landscape of inorganic and organic materials remains largely unexplored, offering a compelling opportunity for computational and experimental discovery of novel multi-element compounds with desirable properties and functionalities. However, the immense scale of the composition-structure-property space to be explore makes exhaustive experimental exploration impractical. Recent advances in artificial intelligence (AI) and machine learning (ML) technologies offer transformative strategies for efficiently navigating these complex spaces to significantly accelerate the pace of materials discovery [1-13].

Most of the current AI/ML approaches for materials discovery are based on the structure types or motifs from known structure databases. Despite the vast array of crystal structures cataloged in current materials databases, many undiscovered materials with superior properties and functionality can lie outside the established catalog. AI/ML approaches that rely on known structure types in the databases do not reveal new structure types beyond those that are known. Empirical rules can also be used to favor specific motifs and discover new compounds and structures. For example, the antagonistic-pair rule [14] has been used to discover new compounds with low dimensional structures that evolve from the topological need to separate two mimicable elements by a third element in the solid state. For example, a recent experimentally synthesized compound, $La_4Co_4Pb$ [15, 16], was missed in previous ML searches [10] based on crystal graph convolutional neural network (CGCNN) [17], because the structure type of the new, pseudo-two-dimensional, $La_4Co_4Pb$ compound was not present in the known structure databases.

To fully map the composition-structure-energy landscapes (i.e., convex hulls) of multi-element systems, it is necessary to develop methods for discovering new structures that can overcome the limitations of existing AI/ML approaches. Although generative AI can theoretically design new structures, its practical application for complex materials remains limited. On the other hand, several computational crystal structure prediction methods based on particle swarm optimization (PSO) [18], or genetic algorithm (GA) [19-21] have been developed for decades. These methods have been shown to be capable of predicting new crystal structures using solely chemical composition information. A major bottleneck for the application of such methods to search for complex multi-element compounds with large unit cells is the lack of accurate and efficient interatomic potentials. Although quantum mechanical calculations based on density functional theory (DFT) can provide accurate description of interatomic interactions for most of the materials, PSO or GA based on DFT calculations are computationally too expensive for complex compounds containing large numbers of atoms. On the other hand, whereas empirical classical potentials can facilitate fast GA searches, the accuracy of the results from such GA searches are questionable. For complex ternary systems, empirical classical interatomic potentials are largely unavailable. Recent advances in AI/ML, especially deep learning based artificial neural networks (ANN), offer a breakthrough to develop accurate and efficient ML interatomic potentials enabling reliable GA searches for the crystal structures of complex compounds.

In this paper, we integrate GA with ANN-ML interatomic potentials and develop a multi-minima iterative genetic algorithm (MMIGA) scheme to efficiently search for energetically favorable crystal structures of complex compounds based solely on the chemical composition information (i.e. formula units and number of formula units per unit cell). The complex ternary La-Co-Pb system is used to demonstrate the performance of the scheme. We show that the ML-enhanced iterative GA method not only correctly captures the structures of the $La_4Co_4Pb$ compound which was missed by CGCNN approach but also predicted a novel $La_5CoPb_2$ compound whose x-ray diffraction well match with the experiment [10].

## 2. ML-enhanced iterative multi-minima genetic algorithm

The flowchart of the ML-enhanced MMIGA for crystal structure prediction is illustrated in **Fig. 1(a)**. Conventional GA is the backbone of the scheme as shown in blue color, while the structure relaxation and potential energy calculation in the GA loop is performed using an ANN-ML interatomic potential as indicated by the golden box. In conventional GA search, when the potential energy of the lowest-energy structure in the pool is unchanged for a certain number of consecutive generations (say, 500), the GA search is considered as converged. The GA search in such a conventional scheme can be trapped in a local minimum. To boost the efficiency of GA for global structure exploration, we added a multi-minima iterative search scheme to the conventional GA as indicated by the green boxes. After each "converged" GA search, a new GA cycle is repeated but with a penalty applied to the structures with similarity to the lowest-energy structures from the previous GA cycles. Such iterative GA cycle $i$ can be repeated from 5-10 cycles or more depending on the complexity of the system. We refer to each GA cycle $i$ as iteration $i$. The idea of adding penalty is analogous to that in metadynamics for rare event simulations [22] to prevent unwanted revisiting the known low-energy states, thus avoiding trapping GA search in a local minimum.

In our GA search, the shape of unit cell, i.e., the lattice constants, is an optimization variable. The lattice constants of each structure in the initial GA pool are generated randomly while keeping the volume of unit cell approximately equal to the sum of total volume of the atoms (assumed to be spheres) in one unit cell. After generations of evolution, we found the lattice constants in the GA pool usually have a narrow distribution, suggesting that the offspring generated for the next generation are likely to have similar unit cell shapes thus trapping the GA search in a local minimum. Therefore, using lattice constants as the collective variables for the penalty energy is convenient to enhance the efficiency of global crystal structure prediction, although other collective variables can also be employed. For simplicity, we use the length of the shortest of the three lattice constants as the collective variable which is denoted as **a**. The penalty energy is modeled by a Gaussian function centered at $a$ as $E_p = \sum_{i=1}^{N} P_i = \sum_{i=1}^{N} E_0 \exp\left[\frac{-(a-a_i')^2}{2\sigma_i^2}\right]$, where $a_i'$ is the shortest lattice constant of the lowest-energy structure obtained from the $i$th MMIGA search. $E_0$ is about 200 meV/atom and $\sigma_i$ is about 1%~5% of the shortest lattice constant of the local minimum structure in $i$th iteration of GA search. The concept of applying penalties for revisiting the known local minima to facilitate the multi-minima search is also schematically illustrated in **Fig. 1(b)**. By adding the penalty energy to the structure visiting the previously known local minima (shaded), the GA can escape from the known local minima and greatly enhance the chance to reach a global minimum.

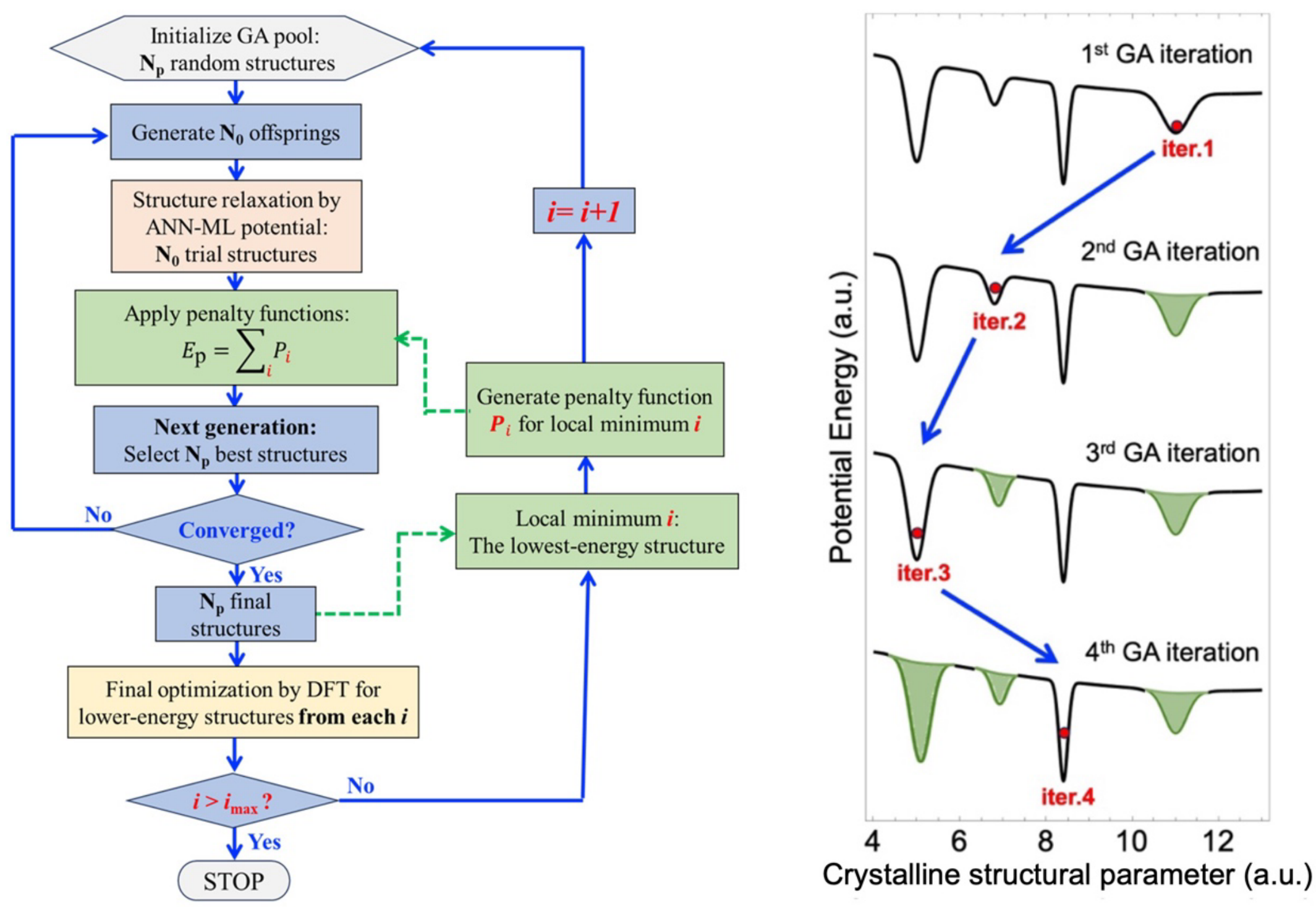

**Fig. 1. (a)** Flowchart of the ML-enhanced multi-minima iterative genetic algorithm (MMIGA) scheme for global crystal structure prediction. The conventional GA loop is plotted in blue. The performance of the GA is enhanced by using ANN-ML interatomic potentials as indicated by the golden box. In the iterative GA cycles as shown in green boxes, a penalty energy $E_p$ is added to the fitness function for the structure selection to the next GA generation pool. **(b)** An illustration of multi-minima GA search through adding penalty energies (shown as green backfilling of the local minima) to avoid revisiting of the known minima.

## 3. Search for new ternary La-Co-Pb compounds

We demonstrated the robust performance of the MMIGA method for global structure prediction of complex compounds using ternary La-Co-Pb compounds containing the immiscible pair of Co-Pb elements as a test case. It was proposed by Canfield [14] that ternary intermetallic compounds formed by adding a third element to a strongly immiscible pair would form with low dimensional motifs associated with the third element either encapsulating (0-D) sheathing (1-D) or layering (2-D) one of the immiscible pair, preventing any direct bonding or contact between them. In almost all cases, experimental efforts to synthesize such phases have produced quasi-low dimensional compounds in support of this hypothesis. Nevertheless, owing to the strong chemical incompatibility between antagonistic pairs, phase spaces containing an immiscible pair of elements remain relatively unexplored, making them promising for novel inherently low dimensional quantum materials discovery.

Several ternary La-Co-Pb compounds (i.e., $La_6Co_{13}Pb$, $La_{12}Co_6Pb$ and $La_5CoPb_3$,) have long been experimentally known [14, 23], each manifesting a clear low dimensional motif. As shown by the black solid circles in **Fig. 2**, while $La_6Co_{13}Pb$ and $La_{12}Co_6Pb$ are thermodynamically stable and lie on the La-Co-Pb ternary convex hull, $La_5CoPb_3$ is about 30 meV/atom above the convex hull according to DFT calculations at T=0K. Using a ML-guided framework powered by crystal graph convolutional neural network (CGCNN) and DFT calculations, Wang et al. predicted two new energetically favorable compounds $La_{18}Co_{28}Pb_3$ and $La_3CoPb$ [10] as shown by the red open circles in **Fig. 2**. The predicted $La_3CoPb$ compound was also subsequently predicted by another ML-guided prediction from Google DeepMind [24]. However, experimental synthesis attempts so far failed to produce the two predicted compounds [15]. Instead, a recent experimental synthesis revealed a new compound $La_4Co_4Pb$ with Z = 4 formula units (f.u.) per cell (i.e., $La_{16}Co_{16}Pb_4$) and a clear 2-dimensional structural motif [15, 16]. The composition of the predicted $La_{18}Co_{28}Pb_3$ is somewhat between this $La_4Co_4Pb$ phase and the known $La_6Co_{13}Pb$ phase as can be seen **Fig. 2**, but has a distinct and different actual crystal structure. DFT calculation showed that the formation energy of the $La_{16}Co_{16}Pb_4$ structure is below the known convex hull, confirming this new structure is thermodynamically stable. This structure was missed by the earlier ML-guided CGCNN+DFT prediction [10, 24] because $La_4Co_4Pb$ adopts a previously unknown structure type and therefore cannot be captured with the database of known structures used in the ML studies. We verified that if this structure-type had been included in the database, it would have been captured by the CGCNN-guided prediction. We also noted that the experimental synthesis attempts for the $La_3CoPb$ phase produced samples with sharp peaks in X-ray diffraction (XRD) patterns, but these peaks are poorly fit to that of the predicted $La_3CoPb$ structure. Moreover, the XRD patterns also could not be satisfactorily indexed to other binary or ternary phases in the La-Co-Pb phase space, hinting at the possibility of yet more undiscovered phases in this system. As such, then, we wanted to see if the limitations in the CGCNN approach can be overcome by new structure search using genetic algorithm.

Due to the intricate chemical bonding in this ternary system, which incorporates lanthanides, *3d* transition metals, and heavy elements, along with a high atom count per unit cell (exceeding 30 atoms by 20%), the energy landscape is expected to be highly complex and contain numerous local minima, presenting a significant challenge for determining the ground state structures computationally. We will show that the ML-enhanced MMGA approach not only correctly captured the missed $La_{16}Co_{16}Pb_4$ phase but also predicts a new phase, $La_5CoPb_2$, which we experimentally confirm and identify as the primary product produced in earlier attempts to grow $La_3CoPb$.

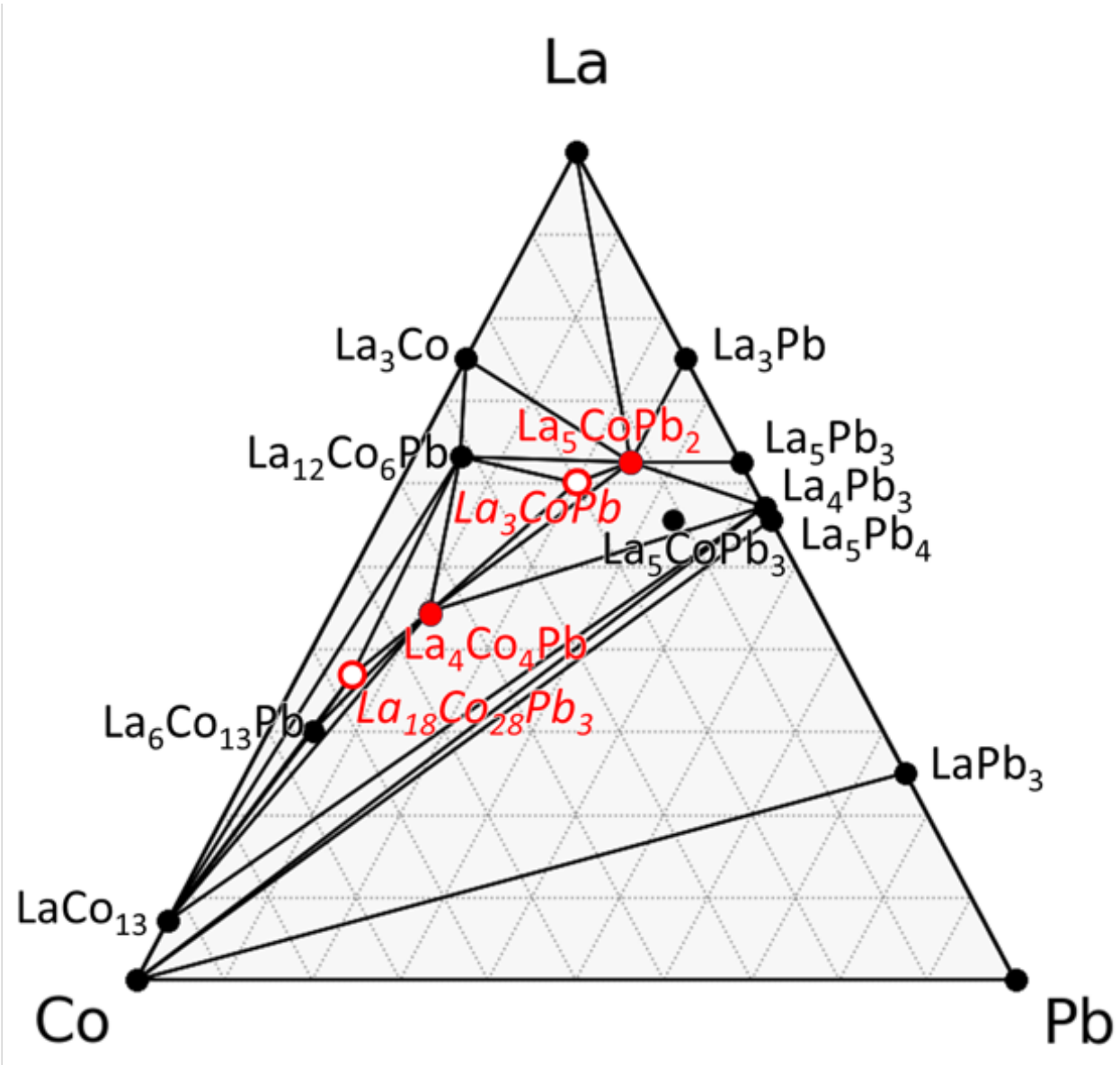

**Fig. 2.** The convex hull of ternary La-Co-Pb system. The known elemental, binary, and ternary phases are shown in black solid circles. Two ternary phases $La_6Co_{13}Pb$ and $La_{12}Co_6Pb$ are on the convex hull, while the third ternary phase $La_5CoPb_3$ is about 30 meV/atom above the convex hull. Two new ternary phases ($La_{18}Co_{28}Pb_3$ and $La_3CoPb$) predicted by ML-guided approach are shown in red open circles. The two phases synthesized by experiment and now revealed by MMIGA are shown in red solid circles.

To enhance the accuracy and efficiency of the GA search, an ANN-ML interatomic potential for La-Co-Pb system was trained using the deep learning software package DeePMD-kit [25-27] and training data from DFT-PBE calculations. Initially, the training dataset included atomistic configurations of pure La, Co, and Pb liquids as well as those at the composition of $La_xCo_yPb_{100-x-y}$, $La_xCo_{100-x}$, $La_xPb_{100-x}$, $Co_xPb_{100-x}$ (x or y = 20, 40, 60, 80). Moreover, randomly distorted crystal structures of the known stable phases of La-Co-Pb system (i.e., La, Co, Pb, $La_3Co$, $LaCo_{13}$, $La_3Pb$, $La_5Pb_3$, $La_4Pb_3$, $La_5Pb_4$, $LaPb_3$, $La_{12}Co_6Pb$, and $La_6Co_{13}Pb$) were also included in the training data set. After initial training, the ANN-ML interatomic potential was then used to perform an initial GA search with the target compositions. New crystalline structures obtained from the initial GA are then used (with random distortions) to refine the ANN-ML interatomic potentials for better accuracy. More details of the ANN-ML interatomic potential development will be published elsewhere.

### 3.1 Search for the structure of $La_4Co_4Pb$ phase

We first searched for ground-state structure of $La_4Co_4Pb$ with 4 formula units, i.e., $La_{16}Co_{16}Pb_4$ (36 atoms per unit cell) based on composition and f.u. information from experimental studies [15, 16]. Using the developed ANN-ML interatomic potential for structure relaxation and energy evaluation as shown in the golden box in **Fig. 1 (a)**, GA was performed with a structure pool of $N_p = 192$ structures. The structures in the initial generation GA pool are randomly generated. The offspring structures in subsequence generations are then created using the “cut-and-paste” operations described in previous publications [20, 21]. At each GA generation, 25% of the structures (i.e. 48 structures) with highest energy in the pool are replaced by the offspring. The penalty energy is applied after the 1st iteration to guide the GA to search for low-energy structures with different lattice constants. When the potential energy of the lowest-energy structure in the pool is unchanged for 500 consecutive GA generations, the GA search is considered “converged”. Then the 12 lowest-energy structures in the final pool of each GA iteration were refined by DFT calculations. The lowest-energy structure from the DFT optimization is taken as the local-minimum structure from the GA search in this iteration. This procedure is repeated for ($i$=12) iterations, ensuring the applied penalty energies span a wide range (4-9 Å) of the collective variable, corresponding to the shortest lattice constant in this study, as shown in detail in the Methods section.

The evolution of the potential energy of the lowest-energy structure in the pool and the average energy of the pool as the function of GA generation in different iterations of the GA are shown in **Fig. S1 (a)-(l)** respectively in Supporting Information. The local-minimum structure obtained from each GA iteration is also shown in the insert respectively. We can see that the local-minimum structures from different iterations of MMIGA search are very different. The top panel of **Fig. 3 (a)** shows the potential energies (per atom) of the structures visited by the GA at each iteration ($i$=1-12) calculated by the ML interatomic potential. More details about the potential energy and penalty energy distributions in different MMIGA iterations can also be found in the method section. The capability of our MMIGA scheme to explore various local minima structure can be clearly seen from the plot; each iteration locates different local minimum basins in the overall energy landscape of the system. By selecting the 12 lowest-energy structures from each iteration of GA, we perform DFT calculations to further relax these structures and obtain more accurate total energies for these structures. In the lower panel of **Fig. 3 (a)**, the cohesive energies (per atom) of the structures from the DFT calculations are also plotted as the function of the collective variable (i.e., shortest lattice constant **a**). We can see that the local-minimum structure from the 7th iteration has the lowest energy, making this the global minimum structure for the $La_4Co_4Pb$ phase. This structure has a *Pbam* space-group symmetry. We also note that there is another structure with space-group symmetry of *Cmc2_1* also from the 7th GA iteration. This energy of this structure is only 4 meV/atom higher than that of the *Pbam* structure. In **Fig. 3 (b)**, we compare the simulated XRDs of the two lowest-energy structures from the GA search with the XRD from experiment. We can see that the XRD of the lowest-energy *Pbam* structure agrees well with experimental result, suggesting the *Pbam* structure from the 7$^{th}$ iteration is the ground-state structure of the $La_4Co_4Pb$ phase. This structure indeed agrees with the solution obtained from SCXRD, showing that our MMIGA approach can overcome limitations with previous searches reliant on known structure types to predict the correct crystal structure using solely composition information.

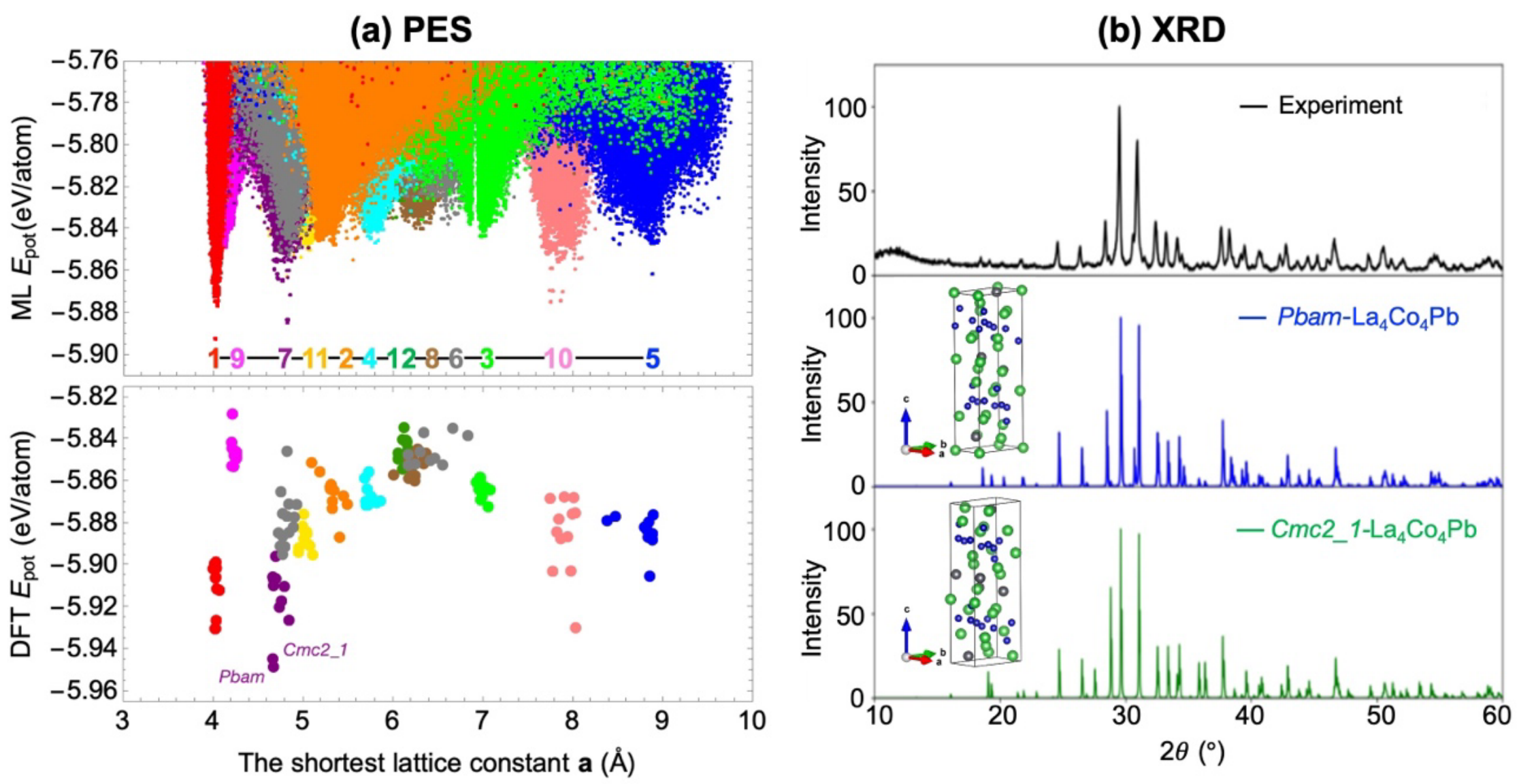

**Fig. 3.** MMIGA results for the $La_4Co_4Pb$ phase with Z = 4 f.u. Top panel in **(a)** shows the potential energies (per atom) calculated by the ANN-ML interatomic potential for the structures visited by the GA at all iterations (*i*=1-12). The color numbers under the potential energies (color dots label the iterations). Bottom panel in **(a)** shows the spin-polarized DFT calculated potential energy as the function of the collective variable (i.e., shortest lattice constant **a**) for every 12 lowest-energy structured selected from each iteration of the GA search. **(b)** The simulated XRD from the two lowest-energy structures from the 7$^{th}$ GA iteration with the shortest lattice constant close to 4.8 Å are compared with experimental data from Ref [15].

We also performed DFT calculations to evaluate the convex hull energies ($E_{hull}$) for the low-energy structures predicted by MMIGA. The $E_{hull}$ are calculated with respect to the known convex hull from Materials Project database [23]. The three phases at the vertices of the Gibbs triangle that enclosed the $La_4Co_4Pb$ phase are $La_6Co_{13}Pb$, $La_{12}Co_6Pb$, and $La_4Pb_3$, respectively. Ferromagnetic (FM) configurations are considered in the $E_{hull}$ calculations. A negative value of $E_{hull}$ indicates that the structure from the GA search has a formation energy below the known convex hull thus can be considered as newly predicted stable structure. Our DFT calculations show that the $E_{hull}$ values for the predicted *Pbam* and *Cmc2_1* structures of the $La_4Co_4Pb$ phase are -25 meV/atom, and -21 meV/atoms, respectively, which are well below the previously known convex hull. If we include our newly predicted stable $La_3CoPb$ and $La_{18}Co_{28}Pb_3$ ternary phases from CGCNN [10] and the $La_5CoPb_2$ phase from MMIGA (see Section 3.2 below), the new $E_{hull}$ values are -11 meV/atom, and -7 meV/atoms, respectively.

By detailed analysis of the two competing structures for the $La_4Co_4Pb$ phase obtained from the MMIGA, we find the corresponding structure motifs are very similar and both conform with the antagonistic pair, low dimensional motif, structural hypothesis [14]. As shown in **Fig. 4**, the difference between the *Pbam* and *Cmc2-1* structure is the subtle arrangement of the Co layer. While Co atoms form a buckling Kagome layer in both structures, the buckling patterns of the Co atoms is different. In the *Pbam* structure, the zig-zag Kagome Co layers are tilted alternately in a period of two layers, whereas the tilting directions are the same for all Kagome layers in the *Cmc2_1* structure. The tilting difference is associated with the different ways the triangles of the Kagome lattice are capped by Co atoms from above or below the layer as shown in the figure. Although half of the triangles in Kagome lattice are capped by Co atoms for both the *Pbam* and *Cmc2_1* structure, the mode of Co atoms capping on Kagome lattice is different. In the *Pbam* structure, Co atoms cap the adjacent (head-to-head) triangle pairs in Kagome lattice (one triangle is from above and the other triangle is from below), where the Co capped triangle pairs are alternately arranged with uncapped triangle pairs in parallel. On the contrary, in the *Cmc2_1* structure, the Co capped triangles are adjacent to the three uncapped triangles, i.e., for each adjacent triangle pair in Kagome lattice only one triangle is capped by Co atom and there is no single triangle pair in which the two triangles are both capped by Co atoms. In addition, for both the *Pbam* and *Cmc2_1* structure, the triangles capped by Co from above and the triangles capped by Co from below are alternately arranged in parallel. Both structures are dynamically stable without imaginary vibrational modes from phonon calculations as can also be seen from the bottom panels of **Fig. 4**.

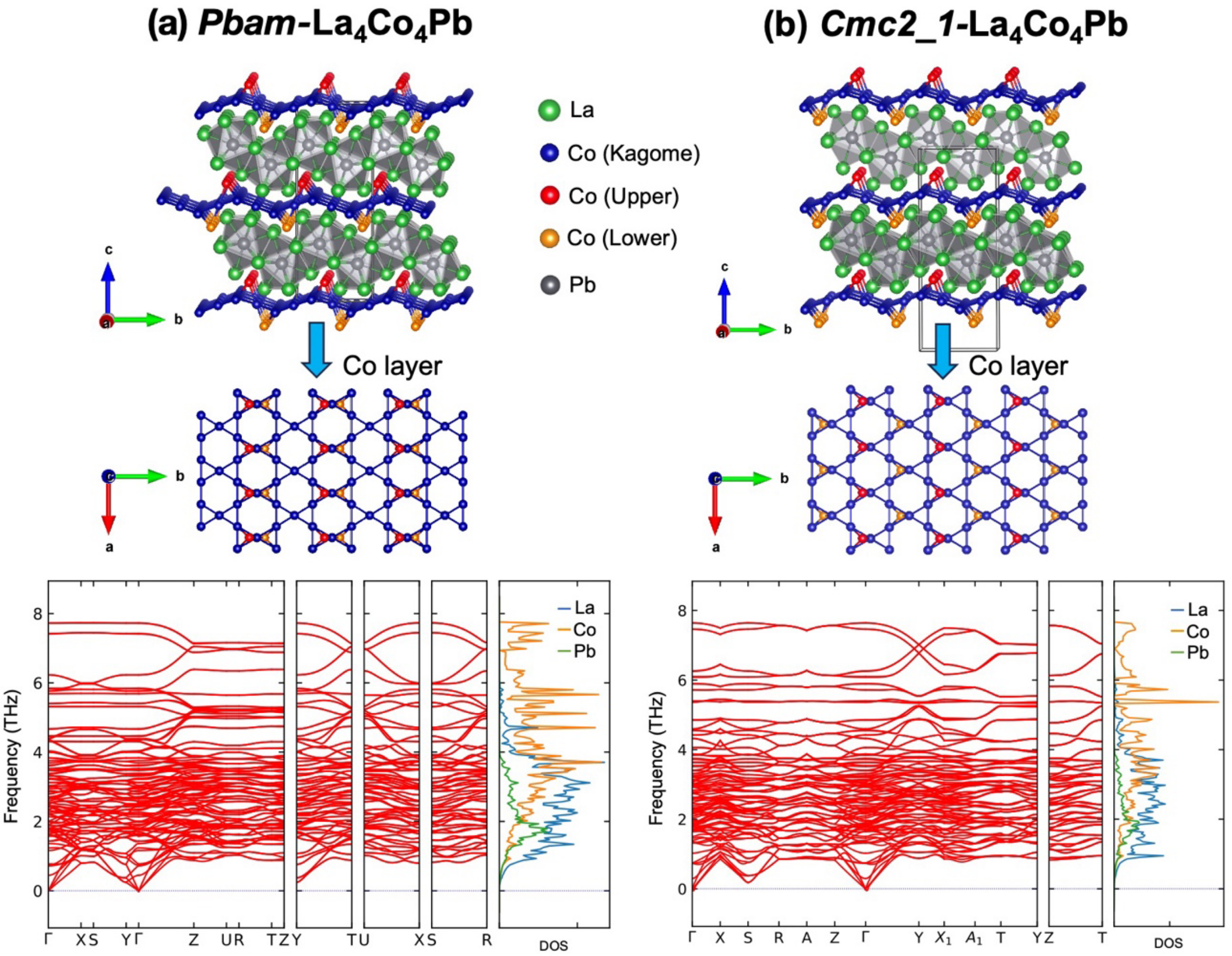

**Fig. 4.** The atomistic structures (top and middle panels), phonon dispersions and density-of-states (bottom panel) for the two competing lowest-energy structures of the $La_4Co_4Pb$ phase: **(a)** the *Pbam* structure from 7th MMIGA iteration of the MMIGA search; **(b)** the *Cmc2_1* structure from the 7th iteration of the MMIGA search.

In addition to the two competing lowest-energy structures discussed above, our study also reveals two metastable structures with *Pmm2* and *C2/m* space group symmetries, respectively, from the 1st iteration of the MMIGA search. The energies of these two structures are both about only 18 meV/atom higher than that of the 7th iteration's ground state structure. More information about these two metastable structures is given in the Supporting Information.

### 3.2 Search for the structures of $La_5CoPb_2$ phase

As discussed in the introduction, we previously predicted a $La_3CoPb$ phase using an ML-guided CGCNN+DFT method [10]. Whereas attempts to synthesize this phase have so far been unsuccessful, powder x-ray diffraction (PXRD) analysis of the products of many of these trials hinted at the possibility of yet another undiscovered compound in the La-Co-Pb system. Subsequent experimental growth efforts (see methods section) successfully isolated rod-like crystals that were found using energy dispersive spectroscopy (EDS), to have a previously unknown composition of $La_5CoPb_2$. PXRD patterns obtained on the rods (**Fig. 5(b)**) match many of the most intense reflections found in the patterns obtained from the products of the $La_3CoPb$ growth attempts, suggesting that $La_5CoPb_2$ was the primary phase obtained from these efforts. It is interesting to note that $La_5CoPb_2$ is close to nominal composition (6-2-2) used in the previous experimental synthesis [10]. Starting from a nominal 6-2-2 composition, it is feasible that experimental synthesis could end with a 5-1-2 phase with some small mixture of La-Co binary phases. Given that we were able to use our MMIGA scheme to arrive at the experimental crystal structure of $La_4Co_4Pb$ using only the chemical composition as input, we challenged our MMIGA scheme to predict the $La_5CoPb_2$ structure independent of experimental x-ray diffraction results.

Based only on the EDS composition, MMIGA searches were performed to search for the low-energy structures of a $La_5CoPb_2$ phase with Z= 2-4 f.u. per unit cell, since the MMIGA searches were performed independent of experimental results and no f.u information was available, at that time, from experiment. The MMIGA set-ups and procedures are the same as those used for the $La_4Co_4Pb$ phase described above. The results from the MMIGA search with 4 f.u. (i.e., 32 atoms per unit cell) are shown in **Fig. 5**. MMIGA searches with Z=2 and 3 f.u. are shown in Supporting Information. From **Fig. 5 (a)**, we can see that several local-minima with different unit-cell shapes are visited in ten successive GA iterations. More details about the potential energy and penalty energy distributions in different MMIGA iterations can also be found in the method section. The energy of the local minimum structures obtained from the $4^{th}$ and $6^{th}$ iterations are found to exhibit low energies by the subsequent DFT calculations as can be seen from bottom panel of **Fig. 5 (a)**. Among these low-energy structures, the one with space-group symmetry of *Pbam* from the $4^{th}$ iteration is found to have the lowest energy of -5.260 eV/atom. Another structure obtained from the $6^{th}$ iteration has a space group symmetry of *Pnma* whose potential energy is $3-9$ meV/atom higher than that of the *Pbam* structure depending on the pseudopotential and exchange-correlation energy functional used in the DFT calculations (see **Table. I**). However, this *Pnma* structure exhibits imaginary frequency branch in the phonon dispersion curves, and the structure can be relaxed into *Pbam* structure after DFT optimization. The $E_{hull}$ values of the two $La_5CoPb_2$ calculated by DFT using the PBE functional and based on the previously known convex from MP [23] are -26 meV/atom and -19 meV/atom, respectively, for the *Pbam* and *Pnma* structures. If the newly predicted stable $La_3CoPb$ and $La_{18}Co_{28}Pb_3$ ternary phases from CGCNN [10] and the $La_4C_4Pb$ phase from MMIGA (see Section 3.1 above) are included in the convex hull calculation, the $E_{hull}$ values will be -9 meV/atom and -2 meV/atom respectively. Comparison of the simulated XRD for these two lower-energy structures with that of experiment is shown in **Fig. 5 (b)**. We find that the XRD from the *Pnma* structure agrees better with the experimental data obtained from crushed powder of $La_5CoPb_2$ rods, suggesting that the experimentally synthesized structure from the minimal 3-1-1 composition and from the 5-1-2 composition is likely to be the *Pnma* structure $La_5CoPb_2$.

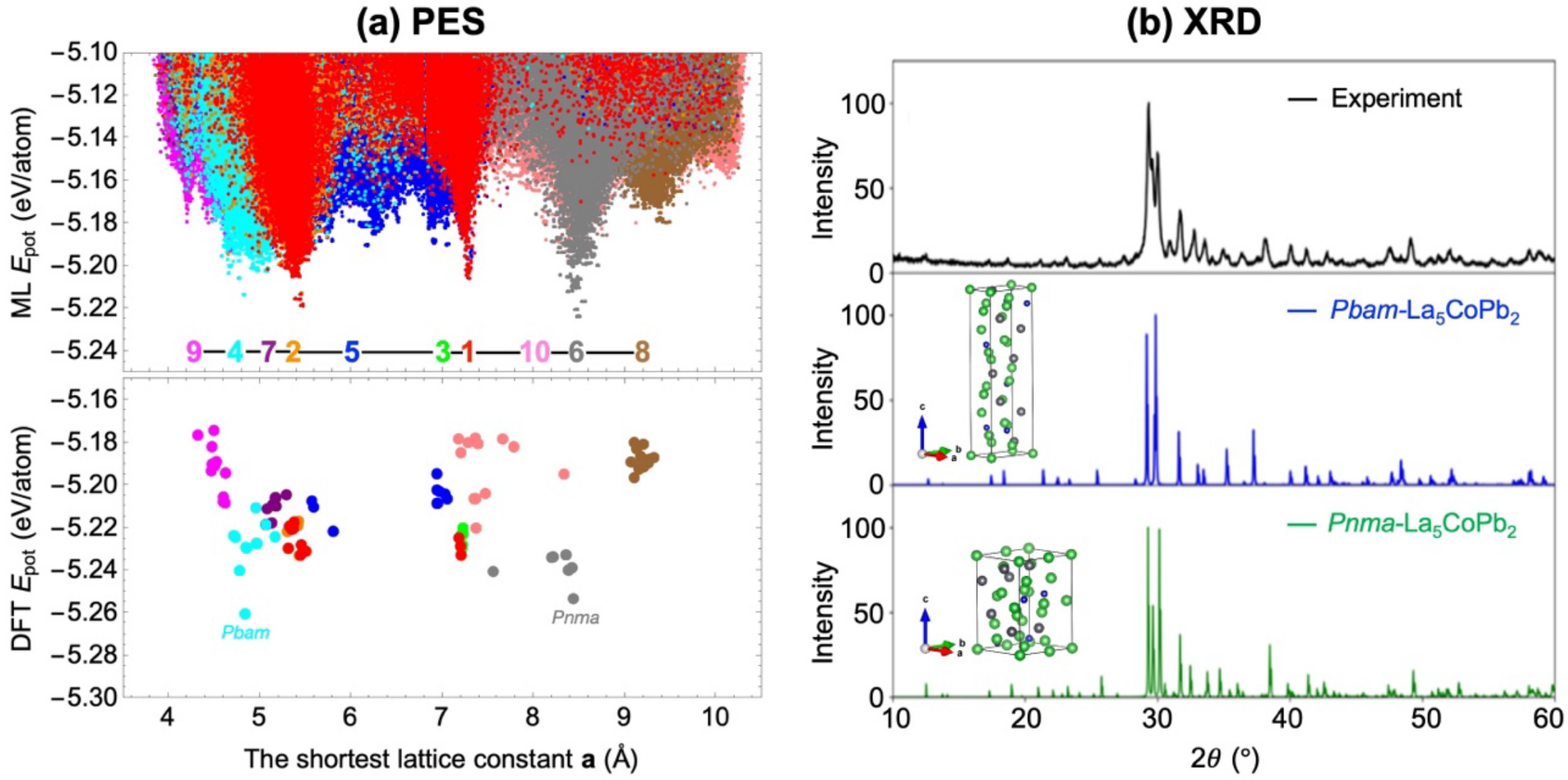

**Fig. 5.** MMIGA results for the $La_5CoPb_2$ phase with Z = 4 f.u. Top panel in **(a)** shows the potential energies (per atom) calculated by the ANN-ML interatomic potential for the structures visited by the GA at all iterations (*i*=1-10). The color numbers under the potential energies (color dots label the iterations). Bottom panel in **(a)** shows the spin-polarized DFT calculated potential energy as the function of the collective variable (i.e., shortest lattice constant **a**) for every 12 lowest-energy structured selected from each iteration of the GA search. **(b)** The simulated XRD from the two lowest-energy structures from the 4$^{th}$ and 6$^{th}$ GA iterations with the shortest lattice constants close to 4.8 Å and 8.4 Å respectively are compared with experimental data from Ref [10].

Detailed analysis shows that the structures of the two competing lowest energy $La_5CoPb_2$ structures obtained from MMIGA are very similar. As shown in **Fig. 6**, both structures are composed of edge-sharing chains of Co-centered Co-La octahedra that extend down the crystallographic *b*–axes in the *Pbam* and *Pmna* structures. Within the *ab*–plain, there are additional chains of La atoms that sit between the $LaCo_6$ octahedral chains. The Pb atoms occupy the centers of voids formed by the La framework, forming a 3D network of tri-capped $PbLa_9$ coordination polyhedra that have face-sharing connectivity along the *a*– and *b*–axes and corner sharing connections along the *c*–axis. It is interesting to note that the salient Pb-La coordination geometry and connectivity is identical to that observed in the Pb-La layers of $La_4Co_4Pb$ (see **Fig. 4**). In $La_5CoPb_2$ however, the Pb-La polyhedra are connected along all three crystallographic axes, forming a 3D framework, whereas in $La_4Co_4Pb$ they are separated into quasi 2D slabs by the Co layers. Both the *Pbam* and *Pmna* structures follow the antagonistic pair hypothesis, manifesting clear separation of the immiscible Co and Pb atoms, producing quasi 1D linear or zig-zag chains of Co that are fully isolated from the Pb by nearest-neighbor coordination to La atoms. We note that the exactly *Pbam* structure is also obtained from the MMIGA search with 2 f.u. as can be seen from Supporting Information.

The main difference between the *Pbam* and *Pnma* structures is that the octahedral chains have linear connectivity along the *b*–axis in the *Pbam* structure whereas the chains are distorted in the the *Pnma* arrangement, extending in a zig-zag pattern down the *b*–axis. Since the La-centered octahedra are connected linearly along the *b*-axis, the primitive unit cell along *b*-axis can be reduced by half, resulting in the same *Pbam* structure with Z = 2. In fact, *Pnma* is a subgroup of *Pbam*, suggesting that the *Pnma* arrangement is just a distortion of the higher symmetry *Pbam* structure. In general, the consequence of structure distortion from a high symmetry 1D chain to a lower-symmetry zigzag chain with a double periodicity should lower the energy. We check this energy difference with different exchange and correlation energy of DFT. The potential energy of the *Pba*m structure is consistently about 3-9 meV/atom lower than that of *Pnm*a structure as shown in **Table I**. The reversed trend observed from the DFT calculation suggests that the DFT may not have sufficient accuracy to resolve the relative stability of the two structures.

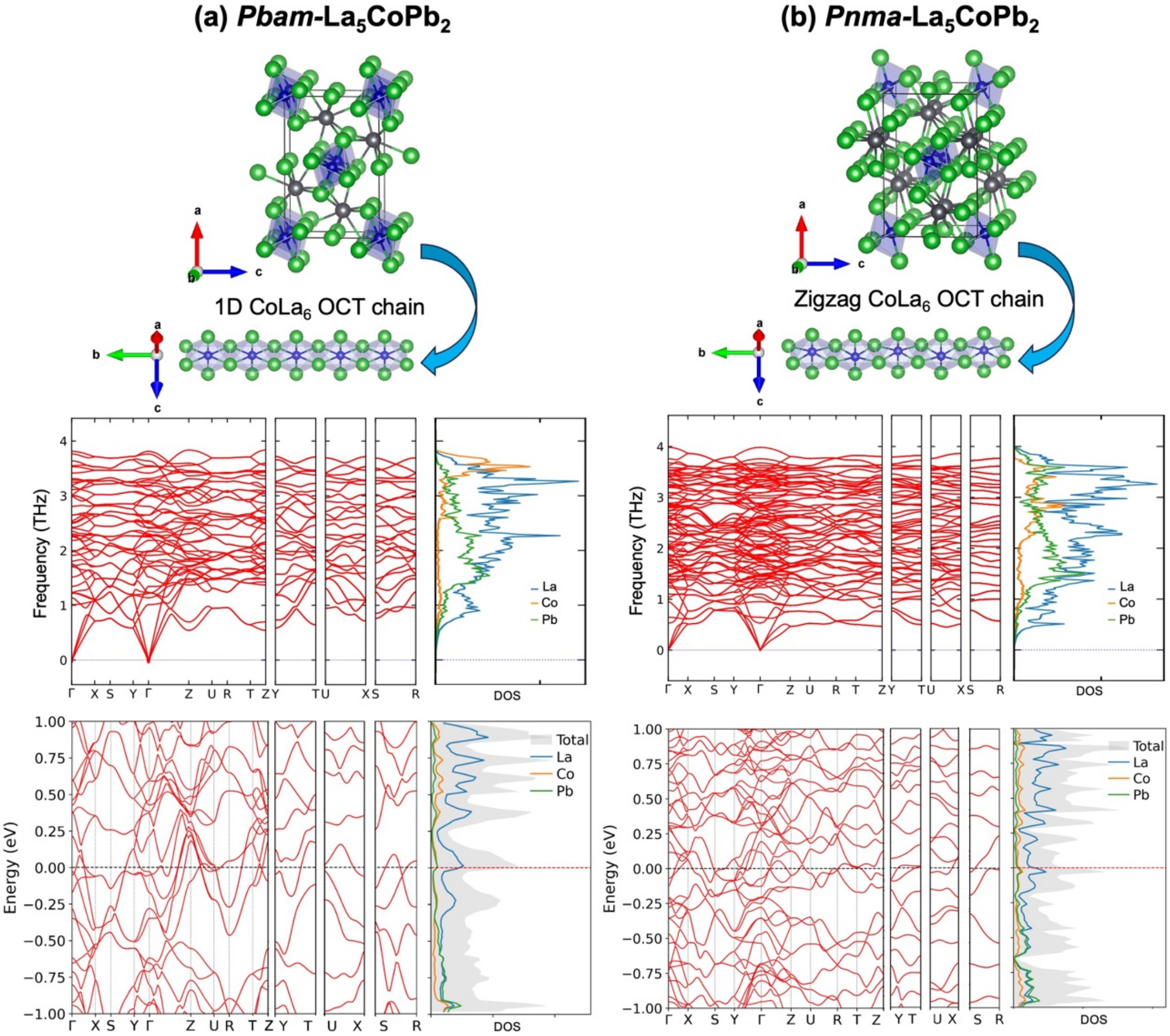

**Fig. 6.** The atomistic structures (upper panel), and phonon dispersions and density-of-states (middle panel) and electronic band structure and density-of-states (lower panel) for the three lowest-energy structures of the $La_5CoPb_2$ phase: **(a)** the *Pbam* structure from 4th iteration; **(b)** the *Pnma* structure from the 6th iteration. In both compounds, the Pb atoms are shelled by La atoms forming Pb-center octahedra which well separate Pb from Co atoms.

**Table. I.** The calculated potential energy differences (meV/atom) between *Pbam*-$La_5CoPb_2$ and *Pnma*-$La_5CoPb_2$ by different exchange and correlation energy of DFT. The potential energy of *Pbam* structure is about $3-9$ meV/atom lower than that of *Pnma* structure.

| | LDA | LDA/Pb_d* | PBE | PBE/Pb_d* | PW91 | PW91/Pb_d* | R2SCAN | R2SCAN/Pb_d* |
|---|---|---|---|---|---|---|---|---|
| *Pbam* | 0 | 0 | 0 | 0 | 0 | 0 | 0 | 0 |
| *Pnma* | 3 | 4 | 7 | 8 | 8 | 9 | 7 | 7 |

* The pseudopotential of Pb is the version of 14 valence electrons which includes *d* electrons.

More details of the crystallography information about these structures will be given in **Tables S1** and **S2** of Supporting Information.

## 4. Experimental synthesis and characterization of the $La_5CoPb_2$ compound

To address whether either of the two low-energy $La_5CoPb_2$ structures predicted by our MMIGA search is correct, we carried out single crystal X-ray diffraction (SCXRD) measurements to experimentally determine the crystal structure. **Table II** lists the atomic coordinates, site occupancy, and thermal displacement parameters determined from the X-ray data.

**Table II.** Atomic coordinates, site occupancies, and isotropic thermal displacement parameters for $La_5CoPb_2$ (space group *Pnma*, $a = 12.8689(3)$, $b = 9.4091(2)$, $c = 8.3496(2)$, $\alpha = \beta = \gamma = 90^o$).

| Atom | site | x | y | z | Occ | $U_{iso}$ |
|---|---|---|---|---|---|---|
| Pb1 | 8d | 0.67166(2) | 0.50497(2) | 0.92496(3) | 1 | 0.01460(8) |
| La1 | 4c | 0.29770(4) | 0.25 | 0.36877(6) | 1 | 0.01705(11) |
| La2 | 4c | 0.70793(4) | 0.25 | 0.65222(6) | 1 | 0.01666(11) |
| La3 | 4c | 0.49765(4) | 0.25 | 0.96133(8) | 1 | 0.01859(11) |
| La4 | 8d | 0.43539(3) | 0.52732(5) | 0.67935(5) | 1 | 0.02100(10) |
| Co1 | 4c | 0.48528(10) | 0.25 | 0.56862(16) | 1 | 0.0186(2) |

We find that $La_5CoPb_2$ adopts *Pnma* symmetry and that the predicted *Pnma* structure discussed above is in excellent agreement with our experimental result. **Fig. 7** shows a comparison of the experimental and predicted *Pnma* structures viewed down each primary crystallographic axis. It is immediately clear that the predicted structure is in close agreement with the experimental result, both being composed of zig-zag chains of $CoLa_6$ octahedra and a 3D network of $PbLa_9$ tri-capped trigonal prisms. In addition, the results of temperature dependent resistance and magnetization measurements are given in **Fig. S10** of Supporting Information.

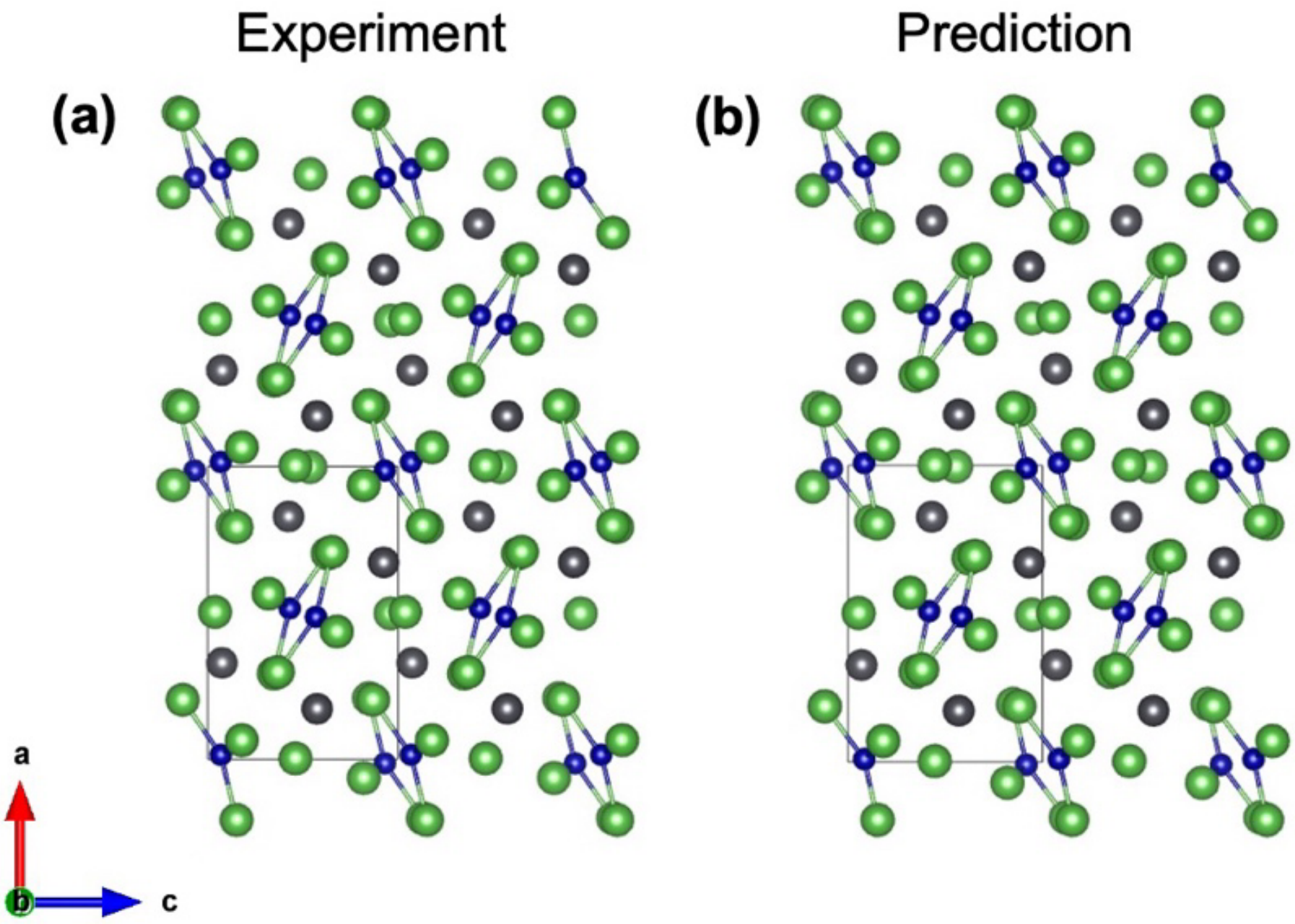

**Fig. 7**: Comparison of the experimental **(a)** and predicted **(b)** crystal structure of *Pnma*-$La_5CoPb_2$ viewed down the crystallographic *b*-axis. The solid black line denotes the unit cell.

## 5. Summary

In summary, we have developed a highly efficient MMIGA framework, driven by ANN-based machine-learning interatomic potentials, to enable practical crystal structure prediction for chemically complex compounds with large unit cells. The key methodological advance is an iterative, metadynamics-inspired strategy in which each "converged" GA cycle is followed by new GA searches that incorporate a penalty against revisiting previously found low-energy basins. Implemented here using a simple collective variable (the shortest lattice constant) in the fitness function, this penalty mechanism reduces premature convergence and promotes exploration of distinct regions of the energy landscape that conventional GA searches can repeatedly miss. The iterative GA search can stop when the penalty energies cover a reasonable range of collective variables. Because the MMIGA workflow relies only on chemical composition as input and does not require pre-enumerated structure prototypes, it bypasses a central limitation of database-trained, motif-dependent ML screening methods when the true structure type lies outside existing catalogs.

We demonstrated the performance of MMIGA on the challenging ternary La-Co-Pb system, where the bonding complexity (lanthanide + *3d* transition metal + heavy element chemistry) and large cell sizes lead to a rugged landscape with many competing local minima. Using the ANN-ML potential to accelerate the GA inner loop and applying DFT refinement only to a small subset of lowest-energy candidates from each iteration, MMIGA successfully recovered the experimentally observed $La_4Co_4Pb$ phase and identified its *Pbam* ground-state structure, addressing the blind spot of earlier CGCNN-guided searches that missed this phase because its structure type was absent from the training database [10]. We then challenged this workflow to predict the crystal structure of another new phase, $La_5CoPb_2$, whose presence was implied from earlier mixed-phase growth attempts in the La-Co-Pb phase space. The MMIGA process predicted two possible structural arrangements, with *Pbam* and *Pnma* symmetry. These predicted structures are very similar, differing only in a subtle distortion of Co-La octohedra. In parallel, we used SCXRD to experimentally show that $La_5CoPb_2$ adopts a *Pnma* structure and that the predicted *Pnma* arrangement is an excellent match with the experimental structure.

It should be noted that while MMIGA with ANN-ML interatomic potential is a vital first step to predict undiscovered, complex compounds, it relies on foundational chemical composition data that remains difficult to determine without experimental guidance. Further development in training ML models to accurately predict compound formation energies based on their composition can serve to streamline composition selection and accelerate the MMIGA for novel materials discovery.

## Methods

### *MMIGA*

In addition to the integration of ANN-ML interatomic potentials, the most notable novel feature of the MMIGA is incorporation of the penalty energy to ensure efficient exploration of the potential energy surfaces of complex systems by GA search.

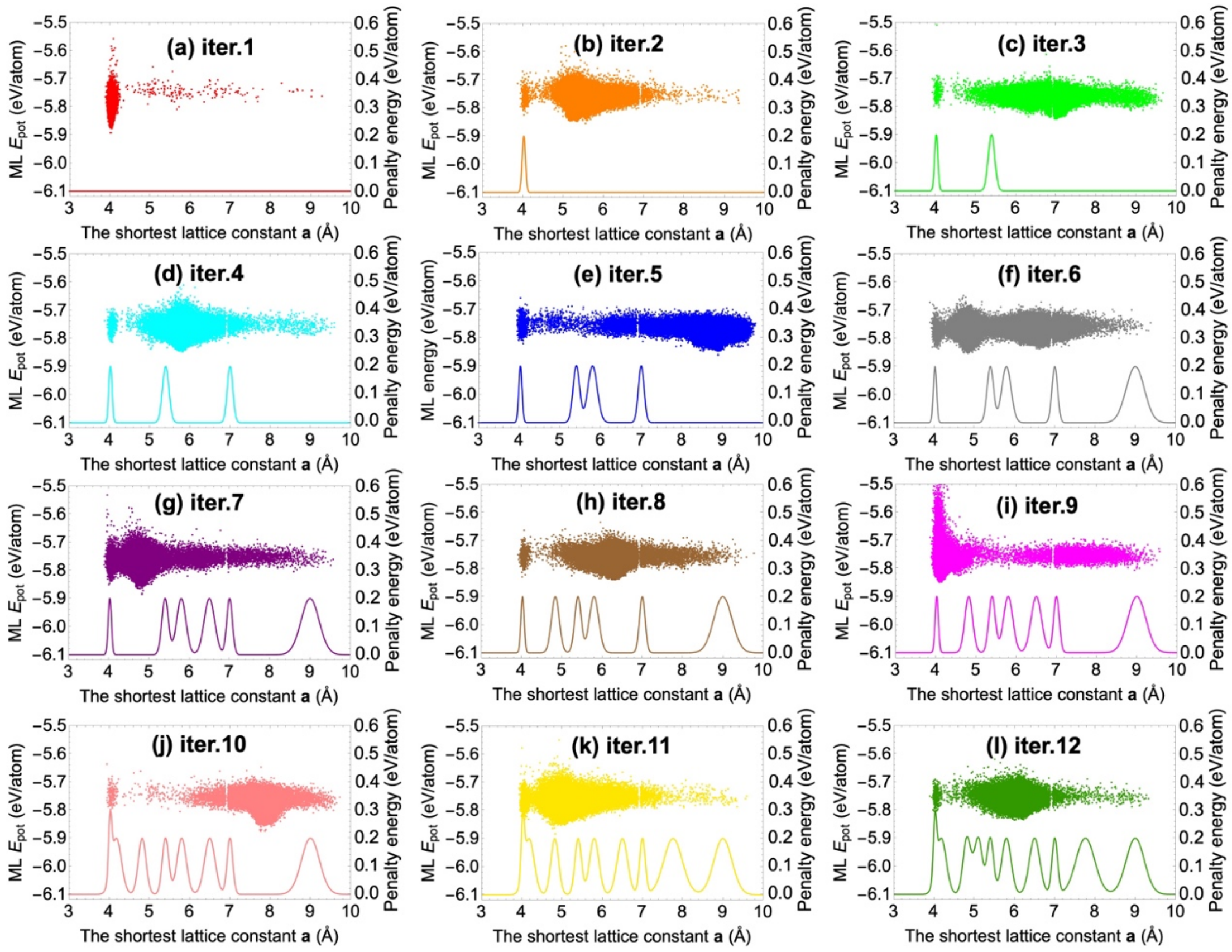

**Fig. 8. (a)-(l)** The ML potential energy distribution (color dots) of all the structures in the GA pools from all generations and the corresponding penalty energies (color lines) in each iteration of the MMIGA for $La_{16}Co_{16}Pb_4$.

The potential energies and the penalty energies sampled during the MMIGA search for the $La_{16}Co_{16}Pb_4$ compounds in the 12 iterations are shown in **Fig. 8 (a)-(l)**. In each plot, the potential energies (left vertical axis) of the structure sampled by GA and calculated by ANN-ML interatomic potential are shown as color dots in the top of the plot, while the corresponding penalty energies (right vertical axis) in each iteration are shown as color lines in bottom of the plot. The horizontal axis indicates the shortest lattice constant, i.e., the collect variable for the penalty energy. In the $1^{st}$ MMIGA iteration, no penalty energy was added and the GA energy distribution concentrated on the $1^{st}$ local minimum with collective variable about 4Å, as shown in **Fig. 8(a)**. In the $2^{nd}$ iteration, the penalty energy of Gauss function is added to the $1^{st}$ minimum as shown in **Fig. 8(b)**. In this iteration, the GA search is hoping out the $1^{st}$ minimum begins to visit the region with collective variable from 4.5-9.5Å until another local minimum around 5.5Å is located. Another penalty energy Gaussian function is added at 5.5Å and proceeded to the 3rd iteration GA search, and so on. As one can see from **Fig. 8**, every iteration can locate a different local minimum. When the penalty energies cover a reasonable range of collect variables (4-9.5 Å in the present case) as shown in the **Fig. 8(l)**, the iteration can be stopped. Then 12 structures with lowest energy by ANN-ML interatomic potential at each local minimum are selected for further DFT optimization and total energy calculations. Final DFT calculated energies are shown in **Fig. 3(a)** above. The MMIGA search identified two competing ground-state structures from the 7th iteration found the minima (i.e., the *pbam-* and *cmc2_1*-$La_4Co_4Pb$ phases) around 4.8 Å as discussed in **Fig. 3(b)** above. The MMIGA search from the 6th iteration also suggests two low-energy metastable structures as shown in the Supporting Information.

Similarly, the MMIGA search for the $La_{20}Co_4Pb_8$ is shown on **Fig. 9.** We can see that the collect variable in the range of 4-10 Å covered with 10 iterations. Multi local minimums are also found from the MMIGA search, among them, two competing ground-state $La_{20}Co_4Pb_8$ with group symmetries of *Pbam* and *Pnma* are found from the 4th and 6th iteration searches, respectively.

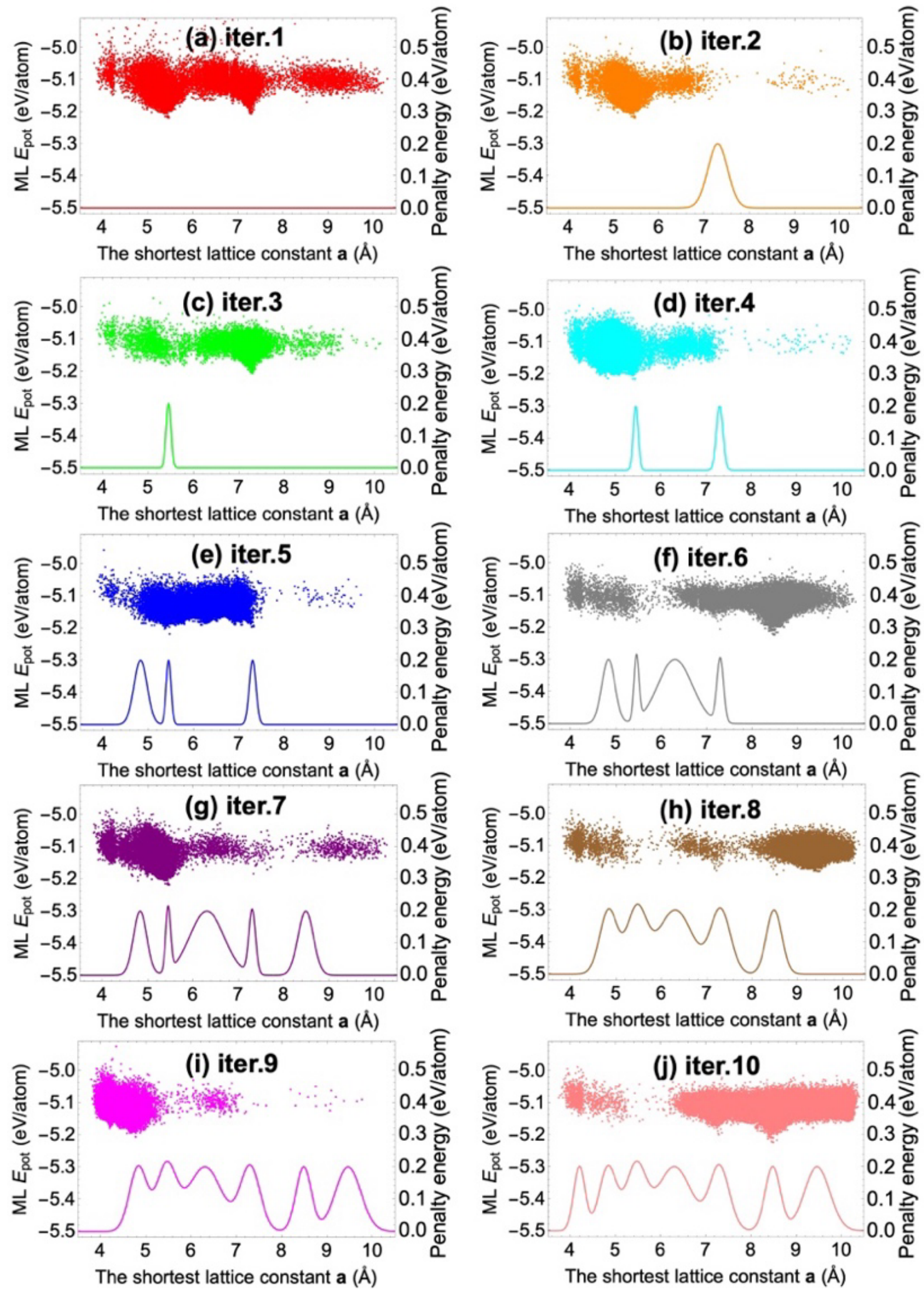

**Fig. 9. (a)-(j)** The ML potential energy distribution (color dots) of all the structures in the GA pools from all generations and the corresponding penalty energies (color lines) in each iteration of the MMIGA for $La_{20}Co_4Pb_8$.

***DFT calculations***

In density functional theory (DFT) calculations [28, 29] of final structure optimization for lowest energy structures, the projector-augmented-wave (PAW) method [30] is used to describe the core-valence electron interaction. The exchange and correlation energy functional on Predew–Burke–Ernzerhof (PBE) form [31] is used unless specified, where the Pb pseudopotential is the version of 3 valence electrons. The cutoff of plane wave basis is set to 520 eV. The Monkhorst-Pack sampling scheme where the k-point grid is $2\pi \times 0.033$ Å$^{-1}$ is adopted for the sampling of the Brillouin zone. The total energy convergence criterion in the electronic self-consistent loop is set to $10^{-5}$ eV. In structure optimization calculation, the crystal is fully relaxed (including the shape and volume of the unit cell and atomic coordinates) until the force on each atom is less than 0.01 eV/Å.

To describe the energetic stability of the structure found in GA searching, the formation energy (calculated by DFT) relative to the known convex hull (denoted as $E_{hull}$) is calculated. The value of $E_{hull}$ for given phase can be calculated by comparing its formation energy with three adjacent known stable phases which are on the vertexes of a triangle (called the Gibbs triangle) in **Fig. 2**. These known phases can be ternary, binary, or elementary crystalline phases. If the $E_{hull}$ of given structure is negative, i.e., if it has formation energy below the convex hull surface, this structure will be energetic stable.

***Phonon calculations***

In phonon calculation, the Phonopy package [32, 33] was used with the finite displacement method. The parameters of DFT calculation are the same as those in structural optimization, except the total energy and force convergence criterion are set to $10^{-9}$ eV and $10^{4}$ eV/Å for high precision relaxation and force constants calculations. In phonon calculation of *Pnma*-$La_5CoPb_2$ phase, the structure is stable (no imaginary vibrational frequencies were found) only using LDA exchange and correlation energy functional. The phonon calculations using other exchange and correlation energy functional (such as PW91 and PBE) show the *Pnma*-$La_5CoPb_2$ phase is unstable. When the structure of *Pnma*-$La_5CoPb_2$ phase is optimized by PW91 or PBE the zigzag $CoLa_6$ OCT chain will relax to the straight 1D $CoLa_6$ OCT chain which is the same as in the *Pbam*-$La_5CoPb_2$ phase. Nevertheless, the XRD data of *Pnma*-$La_5CoPb_2$ phase is closer to the experimental XRD than that of *Pbam*-$La_5CoPb_2$ structure. The potential energy difference between these two competing structures calculated by DFT is very small. **Table. I** shows the calculated potential energies of *Pbam*-$La_5CoPb_2$ and *Pnma*-$La_5CoPb_2$ by different exchange and correlation energy of DFT.

***Crystal growth***

Single crystals of $La_5CoPb_2$ were grown using a self flux method [14]. Elemental La (Ames Lab, 99.9 + %) Co pieces (American Elements, 99.99%), and Pb chunks (Alfa Aesar, 99.99%) were weighed according to a 53:38:7 molar ratio of La:Co:Pb and sealed into a home-made Ta crucible set with Ta caps and frit using an arc melter [14].

The Ta crucible was then flame-sealed in an evacuated fused silica tube and placed in a box furnace. The furnace was heated to 1180°C, held at 1180°C for 8 h, then cooled to 850°C over 100 h, upon which the tube was removed at the excess liquid phase decanted by inverting the tube into a centrifuge with metal rotor and cups. After cooling, the tubes were opened to reveal thin, rod-like crystals.

***Elemental analysis***

The compositions of the crystals were determined by Energy Dispersive Spectroscopy (EDS) using a JEOL NeoScope JCM-7000 Benchtop scanning electron microscope (SEM), using an accelerating voltage of 15 kV. Quantitative analysis of the EDS spectra were done with built-in software SMILE VIEW map with factory standards. We measured multiple spots on each sample to ensure good homogeneity.

***X-ray diffraction***

Powder X-ray diffraction patterns were obtained using a Rigaku Miniflex-II instrument operating with Cu-K $\alpha$ radiation with $\lambda$ = 1.5406 Å ($K_{\alpha1}$) and 1.5443 Å ($K_{\alpha2}$) at 30 kV and 15 mA. Single-crystal X-ray diffraction was performed at 300 K using a Rigaku XtaLab Synergy-S diffractometer with Ag radiation (0.56087 Å) operating at 65 kV and 0.67 mA. The samples were mounted on a nylon loop with vacuum grease. The total number of runs and images was based on the strategy calculation from the program Crysalispro (Rigaku OD, 2023). The data integration and reduction were also performed using crysalispro, and a numerical absorption correction was applied based on Gaussian integration over a face-indexed crystal. The structures were solved by intrinsic phasing using the SHELXT software package and were refined with SHELXL.

**Acknowledgements**

Work at Ames National Laboratory was supported by the U.S. Department of Energy (DOE), Office of Science, Basic Energy Sciences, Materials Science and Engineering Division, including a grant of computer time at the National Energy Research Scientific Computing Center (NERSC) in Berkeley. Ames National Laboratory is operated for the U.S. DOE by Iowa State University under contract # DE-AC02-07CH11358. L. Tang acknowledges the support by the National Natural Science Foundation of China (Grant No. 11304279).

**Data availability**

The data that support the findings of this study are available from the corresponding author upon reasonable request.

**Conflict of interest**

The authors declare no competing interests.

## Supplemental Materials

1. **MMIGA searches for the $La_4Co_4Pb$ (Fig. S1) and $La_5CoPb_2$ (Fig. S2) phases with Z = 4 formula unites**

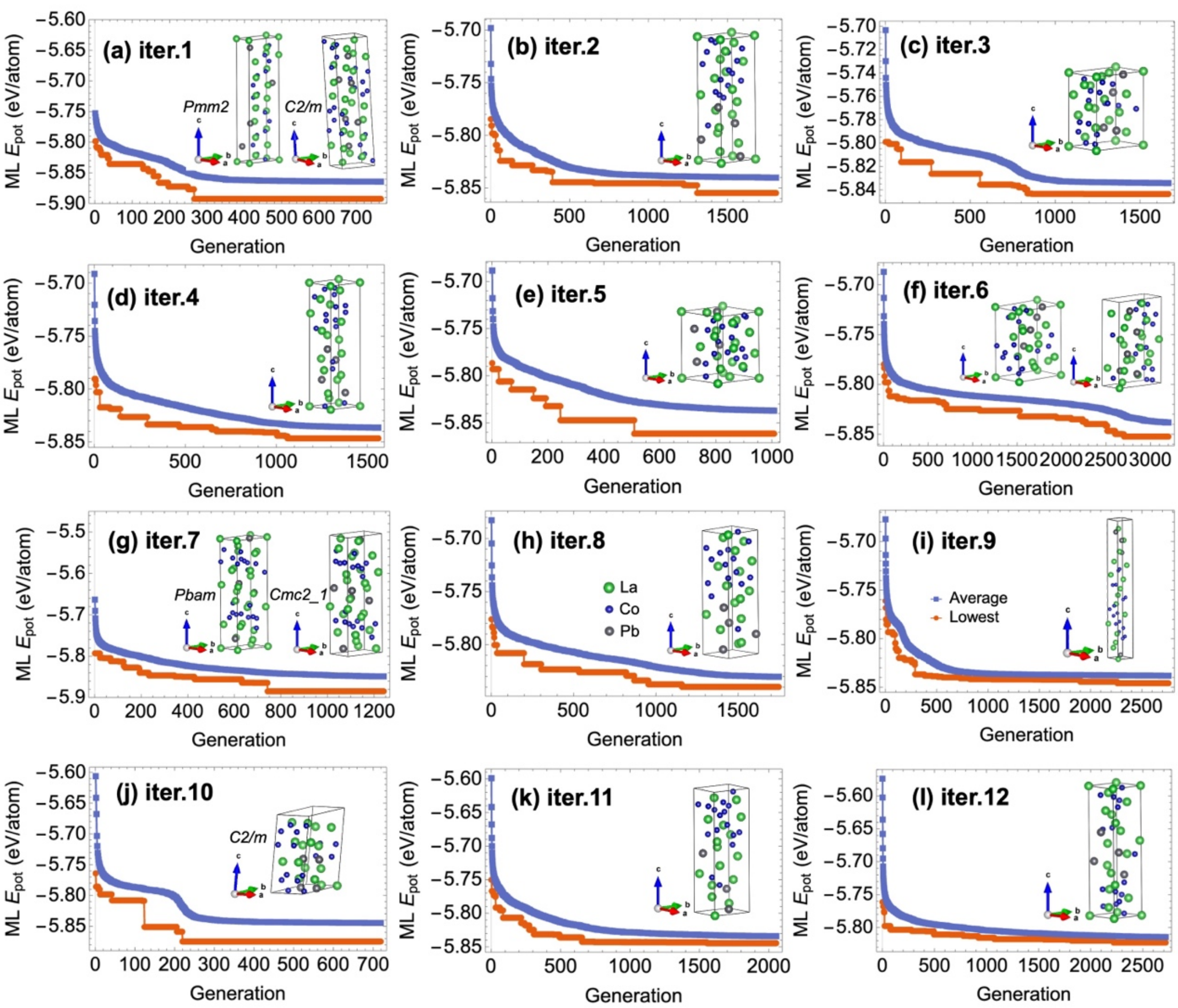

**Fig. S1.** The iterative GA searches for $La_4Co_4Pb$ crystalline phase with Z = 4 formula units (36 atoms/cell). The ML potential energy of the lowest-energy structure in the GA pool and the average energy of the pool as function of GA generations in the 12 successive iterative GA searches are shown in **(a)-(l)**, respectively. The corresponding lowest-energy structures after "converged" GA search in each iteration are also plotted in the insets, with green, blue, and gray balls representing La, Co, and Pb atoms respectively.

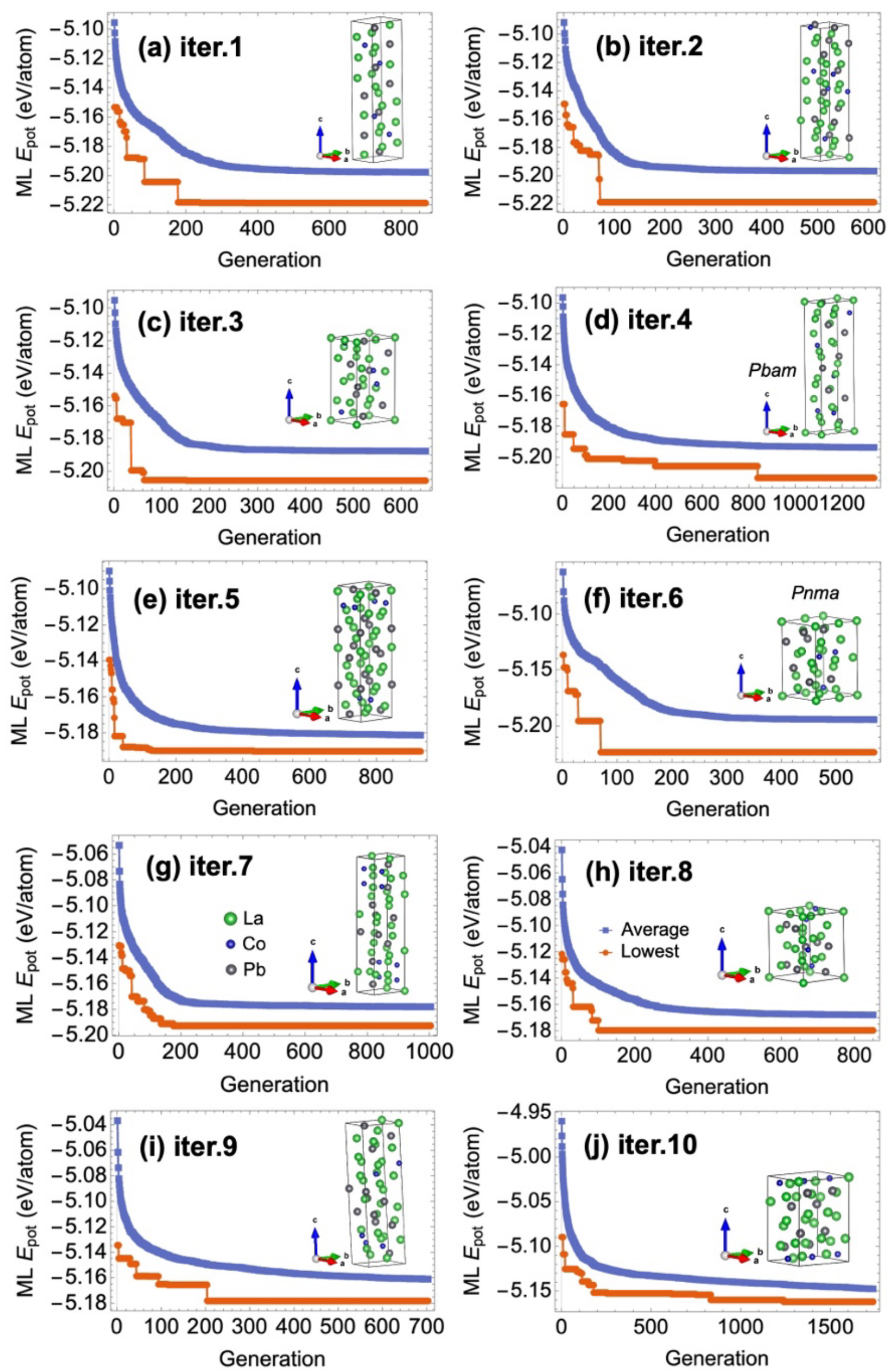

**Fig. S2.** The iterative GA searches for $La_5CoPb_2$ crystalline phase with Z = 4 formula units (32 atoms/cell). The ML potential energy of the lowest-energy structure in the GA pool and the average energy of the pool as function of GA generations in the 10 successive iterative GA searches are shown in **(a)-(j)**, respectively. The corresponding lowest-energy structures after "converged" GA search in each iteration are also plotted in the insets, with green, blue, and gray balls representing La, Co, and Pb atoms respectively.

## 2. Metastable structures of $La_4Co_4Pb$ from MMIGA search

In the 1st iteration of the MMIGA for $La_4Co_4Pb$ (Z = 4 f.u.), our study found two metastable structures with *Pmm2* and *C2/m* space group symmetries respectively. The Co atoms in these two structures also form zigzag buckling layers, but the topology of the atomistic connection is different from the Kagome lattice observed in the ground-state structure. As shown in **Fig. S3**, these two structures exhibit buckled hexagonal-octagonal rings structure for the Co layer. The ground magnetic states of these two structures are both FM and the magnetic moments of the Co atoms are also shown in the figure. No imaginary vibrations are observed for these two structures as can be seen from **Fig. S3**.

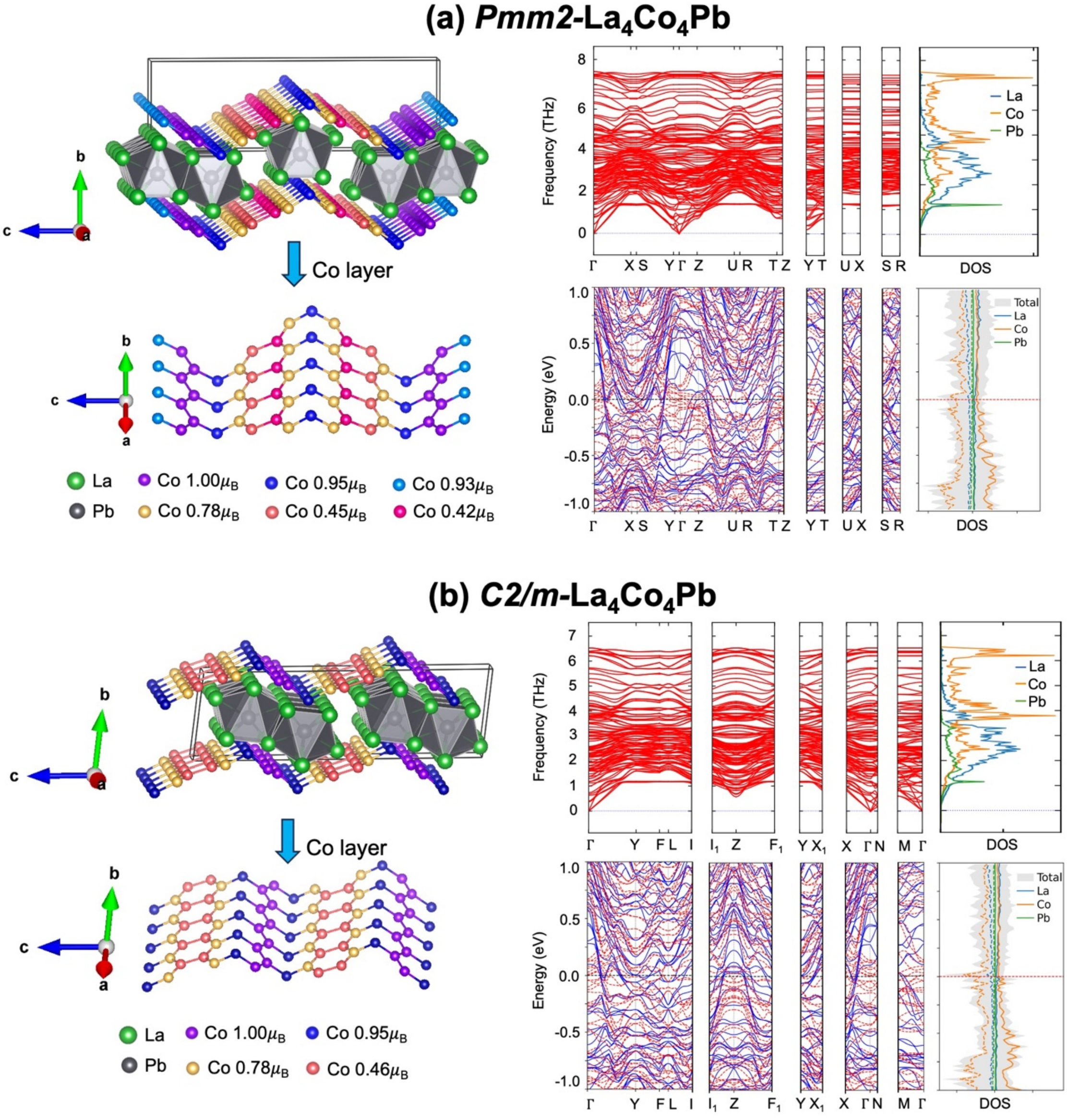

**Fig. S3.** The atomistic structures (left panel), phonon dispersions and density-of-states (right up panel) and electronic band structure and density-of-states (right lower panel) for the two low-energy structures of the $La_4Co_4Pb$ phase for the 1st iteration of MMIGA search: **(a)** the *Pmm2* structure; **(b)** the *C2/m* structure.

### 3. MMIGA searches for the $La_5CoPb_2$ phases with Z = 2 formula unites

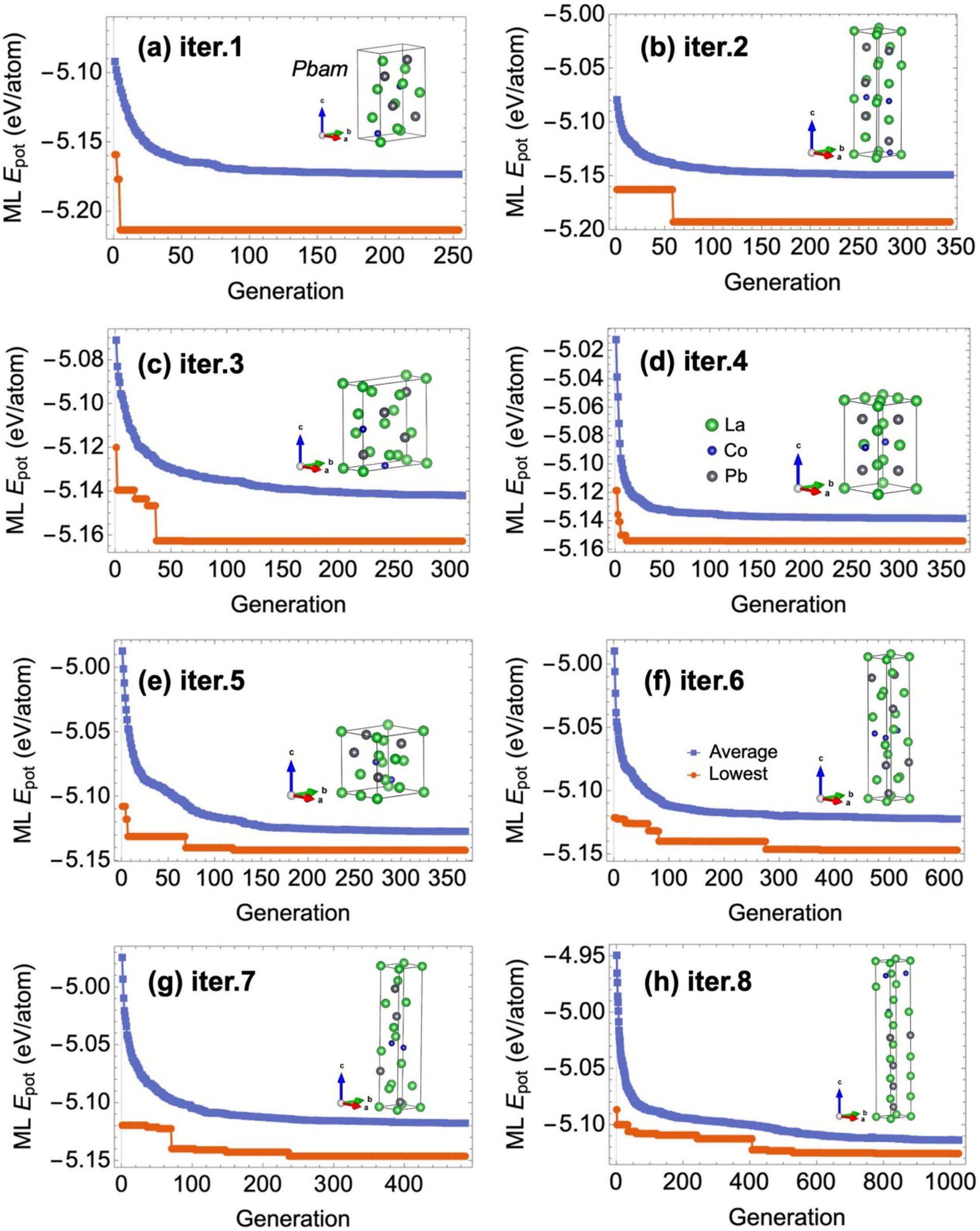

**Fig. S4.** The iterative GA searches for $La_5CoPb_2$ crystalline phase with Z = 2 formula units (16 atoms/cell). The ML potential energy of the lowest-energy structure in the GA pool and the average energy of the pool as function of GA generations in the 8 successive iterative GA searches are shown in **(a)-(h)**, respectively. The corresponding lowest-energy structures after "converged" GA search in each iteration are also plotted in the insets, with green, blue, and gray balls representing La, Co, and Pb atoms respectively.

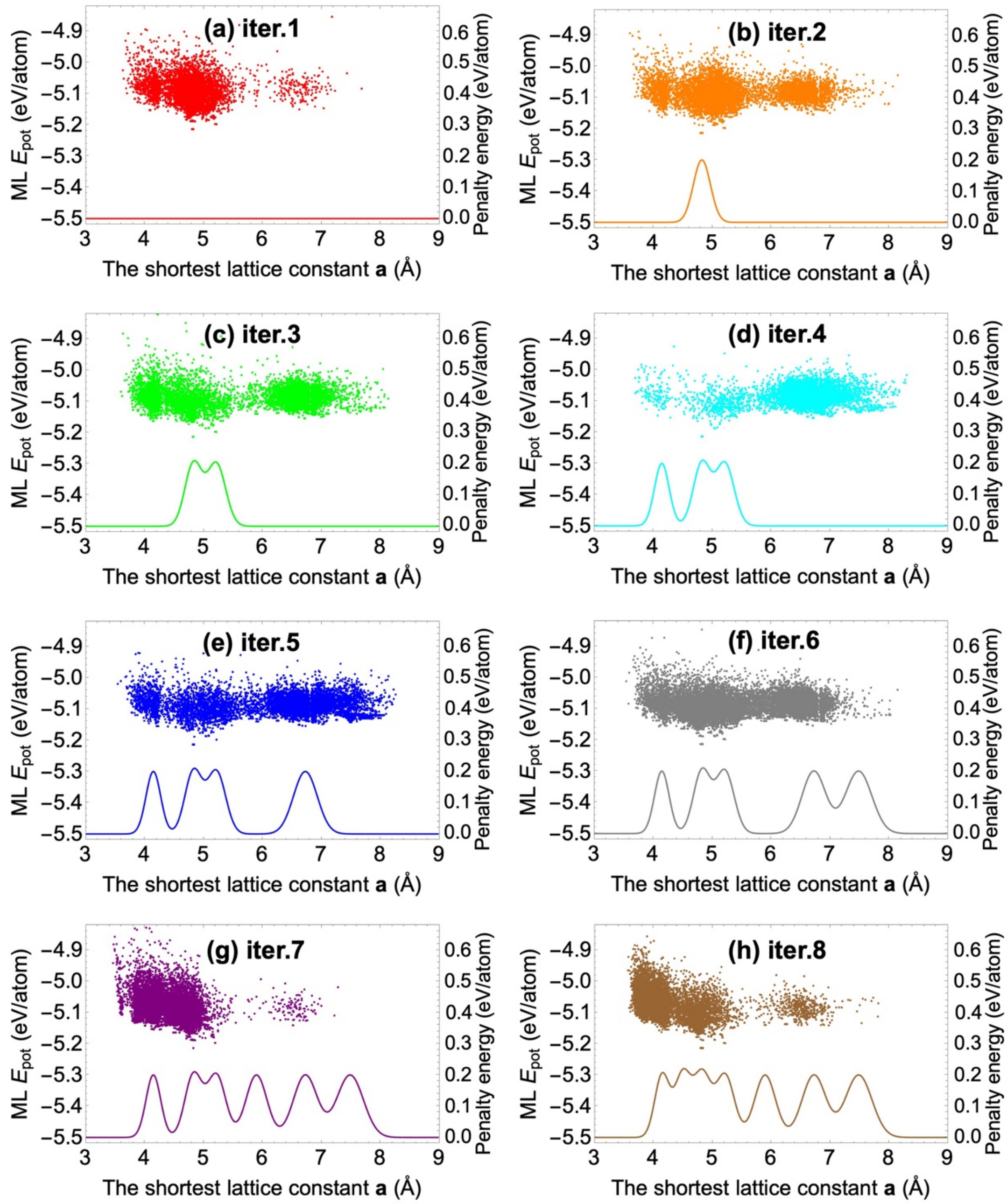

**Fig. S5.** The ML potential energy distribution (color dots) of all the structures in the GA pools from all generations and the corresponding penalty energies (color lines) in each iteration of the MMIGA for $La_5CoPb_2$ (Z = 2 f.u.).

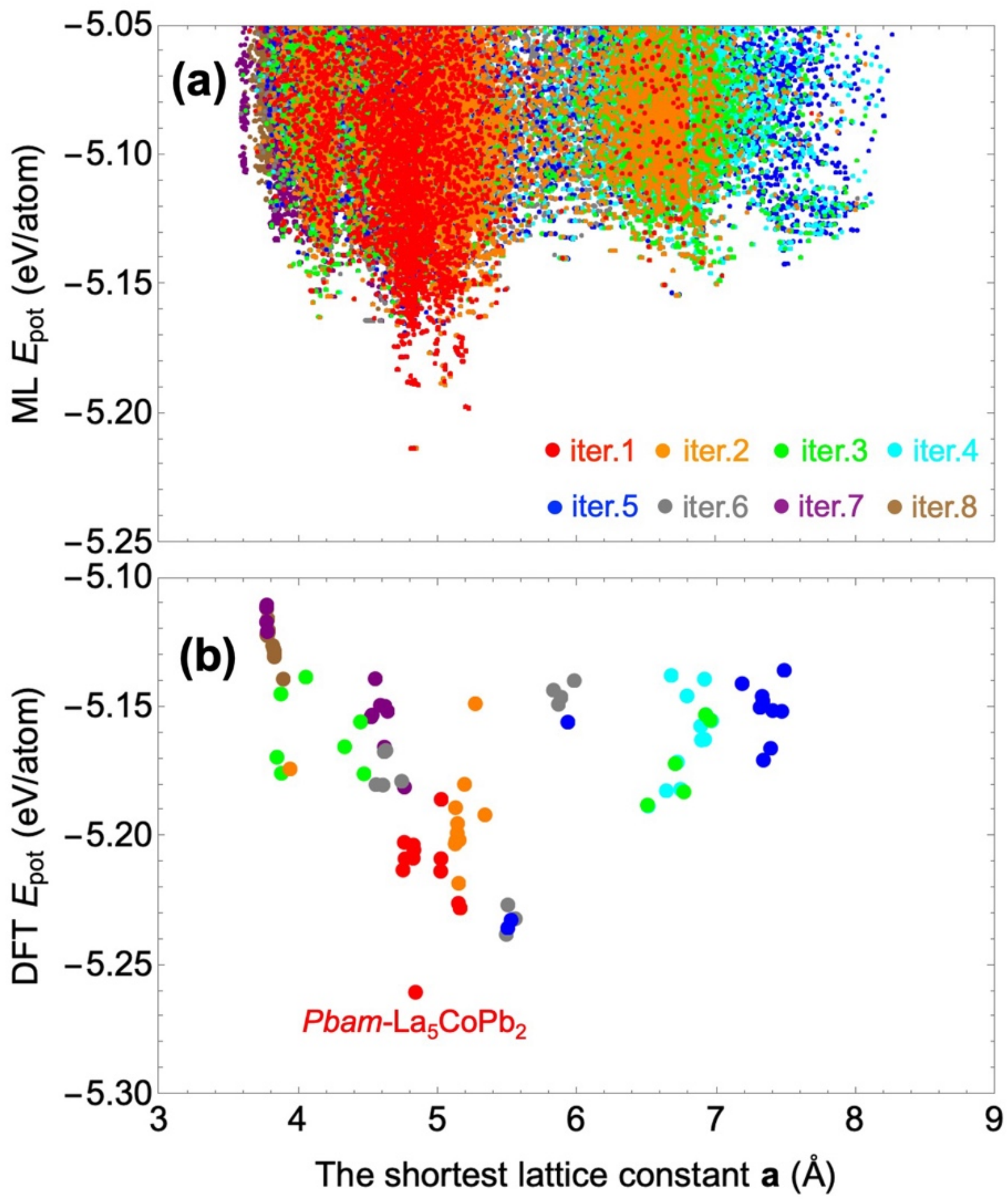

**Fig. S6.** The MMIGA results for the $La_5CoPb_2$ phase with Z = 2 (f.u.). **(a)** The potential energies (per atom) calculated by the ANN-ML interatomic potential for the structures visited by the GA at all iterations (*i*=1-8). **(b)** Spin-polarized DFT calculated potential energy as the function of the collective variable (i.e., shortest lattice constant **a**) for every 12 lowest-energy structured selected from each iteration of the GA search. The 1$^{st}$ MMIGA iteration found the lowest-energy *Pbam*-$La_5CoPb_2$ structure which is also revealed in MMIGA searching for $La_5CoPb_2$ phase with Z = 4 (f.u.).

## 4. MMIGA searches for the $La_5CoPb_2$ phases with Z = 3 formula unites

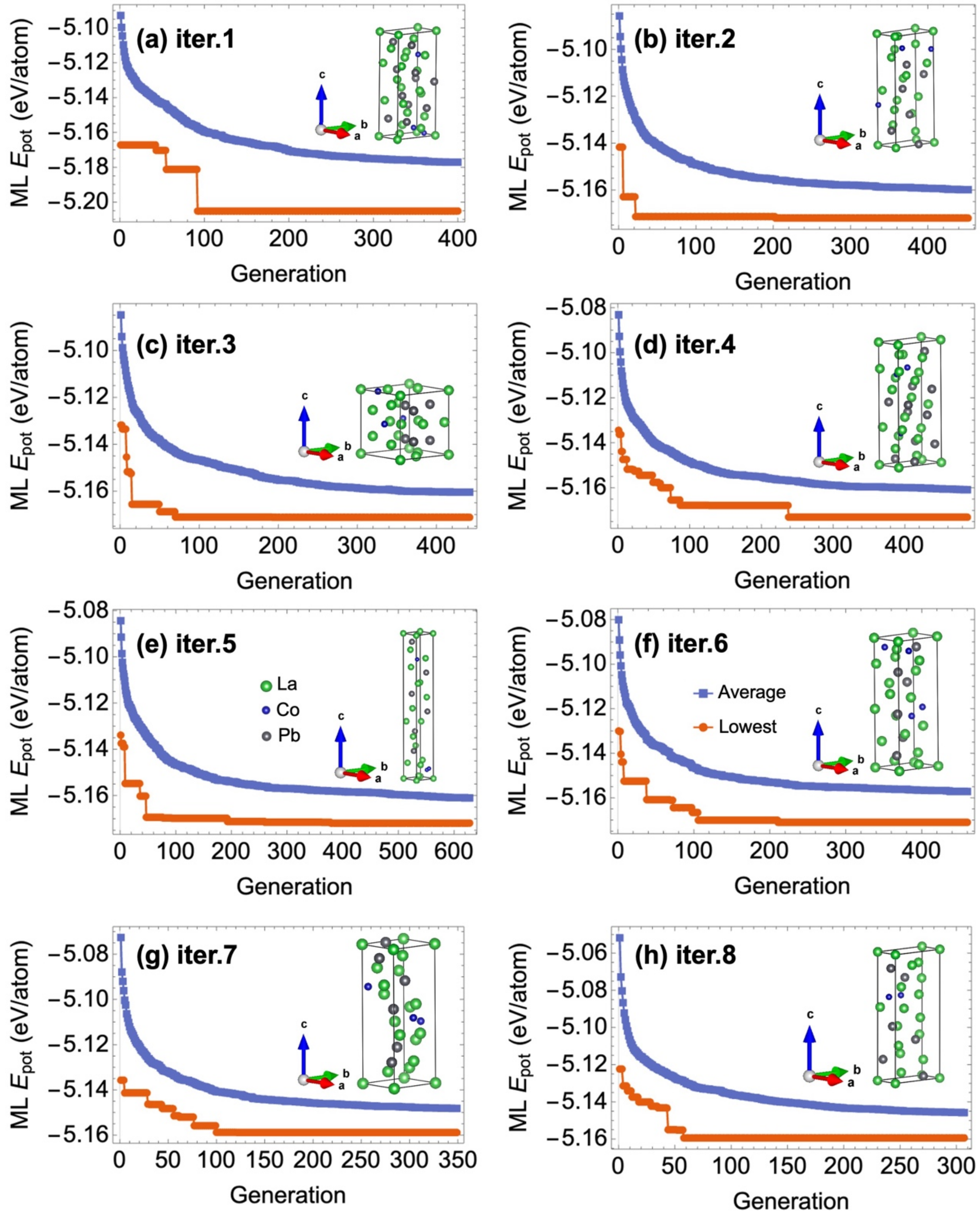

**Fig. S7.** The iterative GA searches for $La_5CoPb_2$ crystalline phase with Z = 3 formula units (24 atoms/cell). The ML potential energy of the lowest-energy structure in the GA pool and the average energy of the pool as function of GA generations in the 8 successive iterative GA searches are shown in **(a)-(h)**, respectively. The corresponding lowest-energy structures after "converged" GA search in each iteration are also plotted in the insets, with green, blue, and gray balls representing La, Co, and Pb atoms respectively.

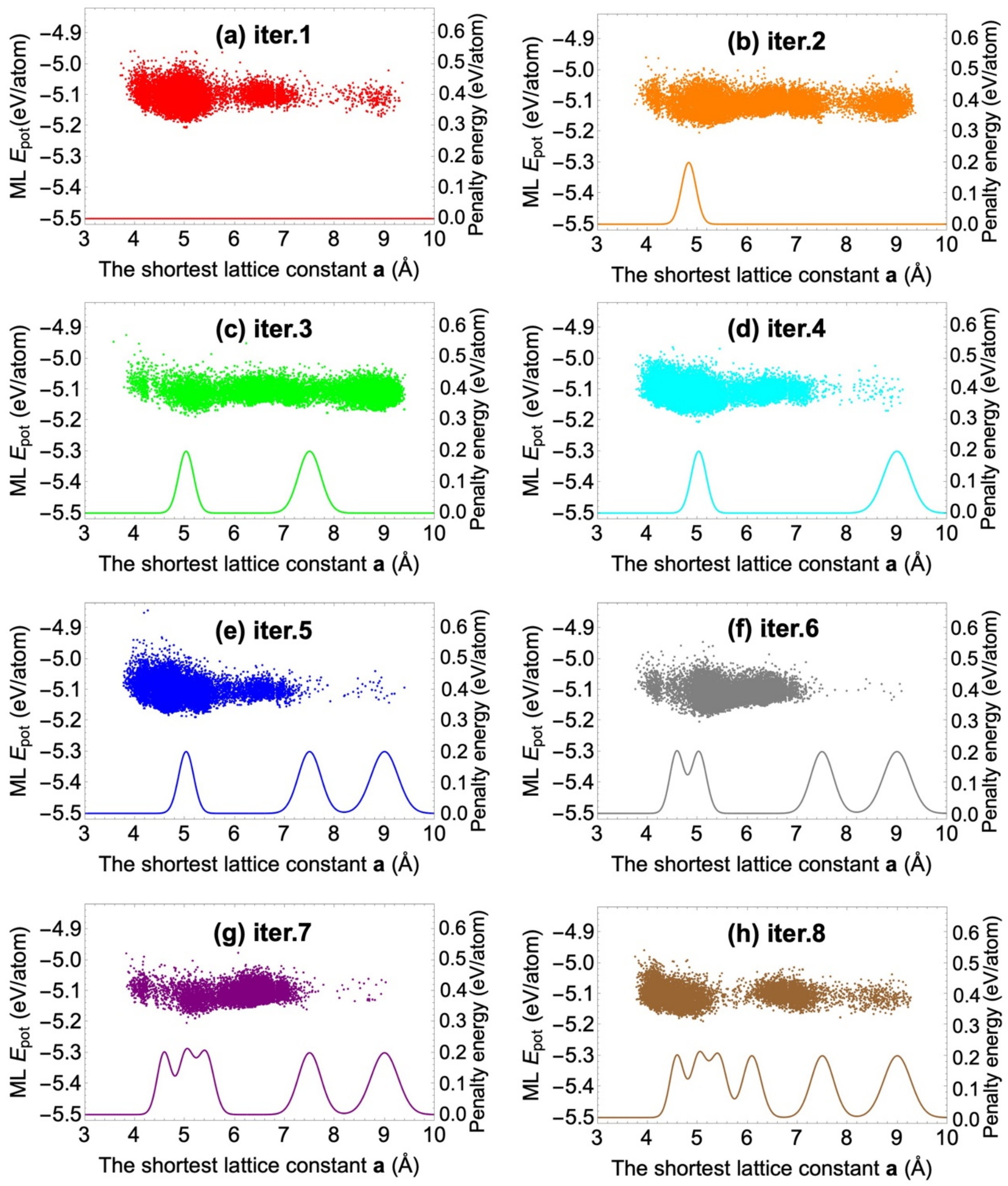

**Fig. S8.** The ML potential energy distribution (color dots) of all the structures in the GA pools from all generations and the corresponding penalty energies (color lines) in each iteration of the MMIGA for $La_5CoPb_2$ (Z = 3 f.u.).

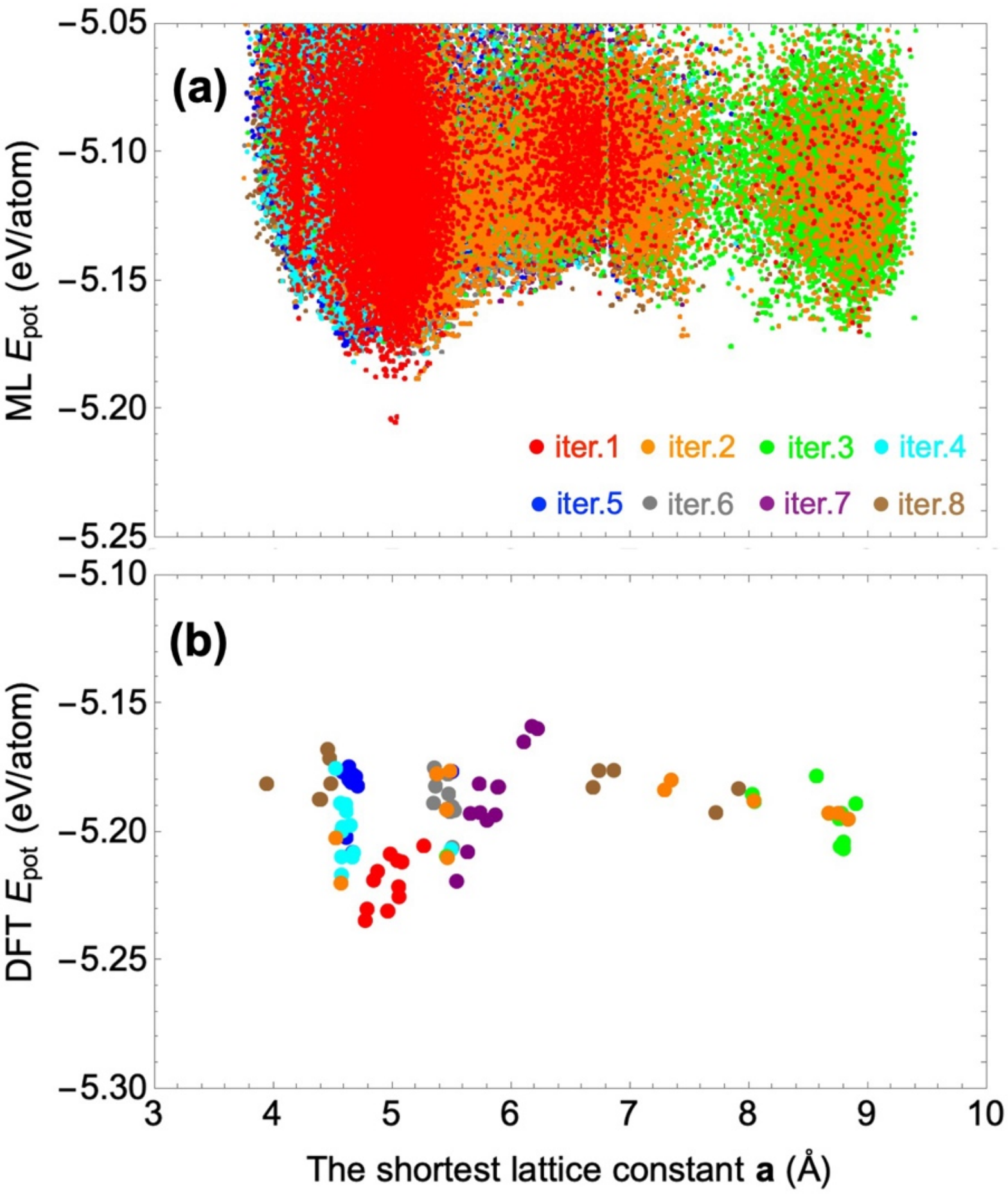

**Fig. S9.** The MMIGA results for the $La_5CoPb_2$ phase with Z = 3 (f.u.). **(a)** The potential energies (per atom) calculated by the ANN-ML interatomic potential for the structures visited by the GA at all iterations (*i*=1-8). **(b)** Spin-polarized DFT calculated potential energy as the function of the collective variable (i.e., shortest lattice constant **a**) for every 12 lowest-energy structured selected from each iteration of the GA search.

## 5. Crystallography information

**Table S1.** Anisotropic thermal displacement parameters for $La_5CoPb_2$

| Atom | $U_{11}$ | $U_{22}$ | $U_{33}$ | $U_{23}$ | $U_{13}$ | $U_{12}$ |
|---|---|---|---|---|---|---|
| Pb1 | 0.01598(13) | 0.01176(12) | 0.01605(12) | -0.00056(5) | 0.00068(6) | -0.00017(5) |
| La1 | 0.0199(2) | 0.01367(19) | 0.0176(2) | 0 | 0.00389(15) | 0 |
| La2 | 0.0204(2) | 0.01315(19) | 0.01642(19) | 0 | 0.00387(15) | 0 |
| La3 | 0.0160(2) | 0.0137(2) | 0.0261(2) | 0 | 0.00195(15) | 0 |
| La4 | 0.01856(17) | 0.02652(19) | 0.01792(16) | -0.00004(13) | -0.00089(11) | -0.00072(13) |
| Co1 | 0.0188(5) | 0.0126(4) | 0.0245(6) | 0 | -0.0067(4) | 0 |

**Table S2.** Single-crystal data and structural refinement information for $La_5CoPb_2$

| Chemical formula | $La_5CoPb_2$ |
|---|---|
| Formula weight (g/mol) | 1167.86 |
| Temperature | 297.15 |
| Wavelength (A, Ag $K_{alpha}$) | 0.56087 |
| Crystal system | orthorhombic |
| Space group | Pnma (# 62) |
| Unit cell dimensions | a = 12.8689(3) Å, $\alpha$ = 90° |
| | b = 9.4091(2) Å, $\beta$ = 90° |
| | c = 8.3496(2) Å, $\gamma$ = 90° |
| Volume ($cm^3$) | 1011.01(4) |
| Z | 4 |
| Calculated density (g / $cm^3$) | 7.673 |
| Absorption coefficient ($mm^{-1}$) | 29.814 |
| Absorption Correction | Face indexed |
| F(000) | 1904 |
| Crystal size (mm) | 0.24 × 0.1 × 0.08 |
| theta range for data collection | 2.295-30.334 |
| Index ranges (min/max, h, k, l) | [-21/22, -16/15, -14/13] |
| Reflections collected | 26007 |
| Independent reflections | 2146 ($R_{int}$ = 0.0632) |
| Completeness | 99.54 |
| Refinement method | Full-matrix least-squares on $F^2$ |
| Data / restrains / parameters | 2996 / 0 / 44 |
| GOF | 1.101 |
| Final R indices [I > 2sigma(I)] | Robs = 0.0397, wRobs = 0.1129 |
| R indices [all data] | Rall = 0.0438, wRall = 0.1226 |
| Extinction coefficient | 0.00024(11) |
| Largest diff. peak and hole ($e \cdot Å^{-3}$) | 5.629 and -4.754 |

## 6. Physical property measurements

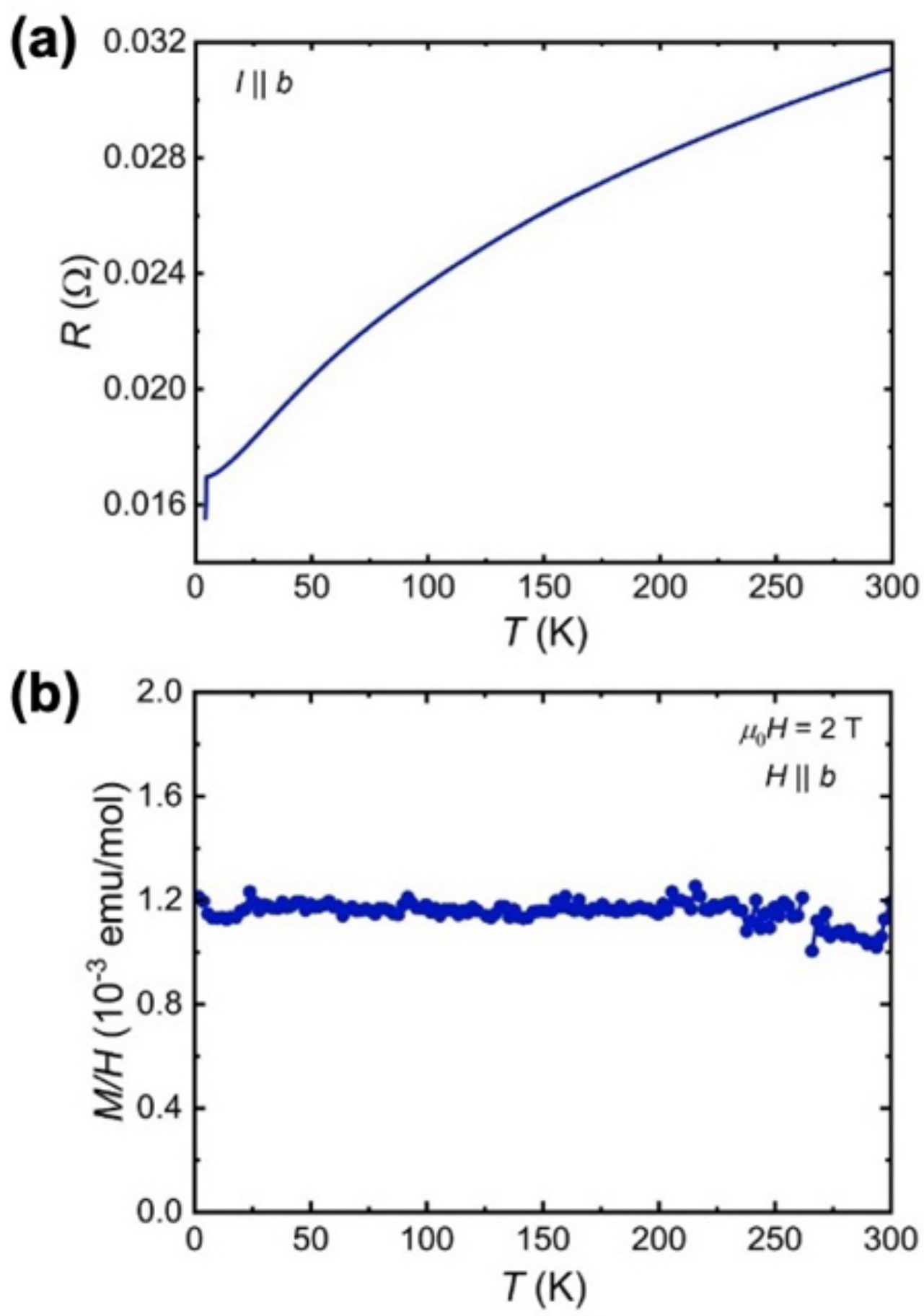

**Fig. S10. (a)** The temperature dependent resistance and **(b)** magnetization (*M/H*) of $La_5CoPb_2$ measured along the *b*-axis (the long axis of the rodlike crystals). Both datasets indicate that the Co atoms are non-moment bearing in $La_5CoPb_2$, showing no signatures of a phase transition and behavior characteristic of a Pauli-paramagnetic metal with metallic $R(T)$ and weak, temperature independent magnetization. A partial superconducting transition is observed at $T_c$ ~ 4.7 K in Figure **(a)**, likely corresponding to a small $La_3Co$ [1] second phase that may form on the surface of the crystals when decanting the excess flux.

Temperature-dependent resistance measurements were performed between 4-300 K in a closed-cyle cryostat (Janis SHI-950) using a Lake Shore AC resistance bridge (model 372) with a frequency of 16.2 Hz and a 3.16 mA excitation current. The temperature was measured with a Cernox 1030 sensor connected to a LakeShore 336 controller. We measured the resistances along the crystallographic *b*-axis (current applied along the long axis of the rods). The contacts were made by spot-welding a 25 μm thick annealed Pt wire onto the the crystals in standard four-point geometry, giving typical contact resistances of ≈1 Ω. A small amount of silver epoxy was painted onto the welded contacts to ensure good mechanical strength.

Magnetization measurements were performed in a Quantum Design Magnetic Property

Measurement System (MPMS classic) SQUID magnetometer operating in the DC mode. The measurements were performed on cleaved samples with both the field applied in-plane and out-of-plane. The samples were mounted on a Kel-F disc, which was then placed inside a straw. The blank disc was first measured at the same temperatures and fields for background subtraction.